\documentclass[twocolumn,english,prb,amsmath,amssymb,eqsecnum,preprintnumbers]{revtex4}
\usepackage[T1]{fontenc}
\usepackage[latin9]{inputenc}
\usepackage{amstext}
\usepackage{amsmath}
\usepackage{amsfonts}
\usepackage{amssymb}
\usepackage{mathtools}
\usepackage{graphicx}
\usepackage{rotating}
\usepackage {bm}
\usepackage[section]{placeins}
\usepackage{color}
\usepackage{multirow}
\usepackage{cellspace}
\makeatletter
\@ifundefined{textcolor}{}
{%
 \definecolor{BLACK}{gray}{0}
 \definecolor{WHITE}{gray}{1}
 \definecolor{RED}{rgb}{1,0,0}
 \definecolor{GREEN}{rgb}{0,1,0}
 \definecolor{BLUE}{rgb}{0,0,1}
 \definecolor{CYAN}{cmyk}{1,0,0,0}
 \definecolor{MAGENTA}{cmyk}{0,1,0,0}
 \definecolor{YELLOW}{cmyk}{0,0,1,0}
}

\def\tr{{\text{tr}}\,}

\def\kF{k_{\text{F}}}
\def\vF{v_{\text{F}}}
\def\NF{N_{\text{F}}}

\def\chis{\chi_{\text{s}}}

\def\epsilonF{\epsilon_{\text{F}}}

\def\sgn{{\text{sgn\,}}}

\def\be{\begin{equation}}
\def\ee{\end{equation}}
\def\bea{\begin{eqnarray}}
\def\eea{\end{eqnarray}}
\def\bse{\begin{subequations}}
\def\ese{\end{subequations}}

\def\chisL{\chi_{\text{s}}^{\text{L}}}

\usepackage{dcolumn}
\usepackage{bm}
\usepackage{amsfonts}
\draft

\makeatother

\usepackage{babel}
\begin{document}
\preprint{arXiv:1812.04592}

\title{Soft Modes and Nonanalyticities in a Clean Dirac Metal}

\author{T. R. Kirkpatrick$^{1}$ and D. Belitz$^{2,3}$}

\affiliation{$^{1}$ Institute for Physical Science and Technology, University of Maryland, College Park, MD 20742, USA\\
                 $^{2}$ Department of Physics and Institute of Theoretical Science, University of Oregon, Eugene, OR 97403, USA\\
                 $^{3}$ Materials Science Institute, University of Oregon, Eugene, OR 97403, USA\\
 }

\date{\today}
\begin{abstract}
We consider the effects of the electron-electron interaction in a clean Dirac metal, i.e., a Dirac system with the chemical
potential not tuned to the Dirac point. We introduce the notion of a Dirac Fermi liquid (DFL) and discuss the soft two-particle 
excitations in such systems, which are qualitatively different from the corresponding excitations in a conventional, or Landau, 
Fermi liquid (LFL). These soft excitations lead to nonanalytic dependencies of observables, including the density of states,
the spin susceptibility, and the specific heat, on the temperature, magnetic field, and wave number, which first appear at 
second order in the interaction. We determine and discuss the nature of these nonanalyticities in a DFL and compare them 
with the corresponding effects in a LFL. Our results are based on very general arguments regarding the existence and 
nature of the soft modes, and corroborated by explicit perturbative calculations. We also discuss their consequences for
 magnetic quantum phase transitions in Dirac metals.
\end{abstract}
%
%
\maketitle

\section{Introduction}
\label{sec:I}

The electronic spectrum of solids in the vicinity of special points in the Brillouin zone where two bands touch
has been the subject of investigations since the early days of solid-state physics.\cite{Herring_1937, Abrikosov_Beneslavskii_1970} 
If one of the touching bands is a conduction band and the other a valence band,
and if the chemical potential is tuned to the touching point (sometimes called a diabolical point), one has the situation of a semimetal, a material
that in a certain sense is in between a metal and an insulator.\cite{Landau_Lifshitz_IX_1991} Even if the
chemical potential lies within the conduction band, the physics that underlies the existence of the touching
point can lead to unusual properties. Much interest in recent years has focused on topological properties
of such materials, which manifest themselves on the surface, and their relation to topological 
insulators.\cite{Wan_et_al_2011, Yang_Lu_Ran_2011, Burkov_Balents_2011} We will focus on a
different aspect, namely, bulk properties such as the density of states and the magnetic susceptibility
for which topological considerations are not important, but which nonetheless are strongly influenced
by the spin-orbit coupling that also leads to the existence of a touching point.

A special case that is of conceptual interest is a system where the
spin-orbit interaction leads to a linear crossing
of doubly degenerate bands, such that the effective Hamiltonian in the vicinity of the crossing point
is reminiscent of 
a massless
Dirac Hamiltonian,\cite{Zhang_et_al_2009} 
although the effects we will discuss persist if the crossing point is gapped out (i.e., if the Dirac
system is massive rather than massless).
We will consider a situation
where the chemical potential lies within the conduction band, so that
he material is a true metal with
an extended Fermi surface (as opposed to a semimetal with a point-like Fermi surface) and well-defined
quasi-particles. In this sense, we will consider a Fermi liquid. 
However, the spin-orbit interaction that leads to the band crossing, whether or not the latter is
gapped out, modifies the properties of the Fermi liquid in ways that make it qualitatively different
from an ordinary Landau Fermi liquid (LFL) that is realized in the absence of a spin-orbit
interaction. This leads us to the notion of a Dirac Fermi liquid (DFL), which differs from a LFL
in the single-particle spectrum, but more importantly in two-particle correlations that are important
for the behavior of various thermodynamic properties, as well as the density of states and
transport coefficients. As we will see, the spectrum of soft two-particle correlations is qualitatively
different in a DFL than in a LFL. Interestingly, this is true even if the spin-orbit interaction is weak
and qualitatively affects the single-particle spectrum only asymptotically close to the Dirac point
(in gapless systems), or not at all (in gapped ones).
By `Dirac Metal' we mean a metal whose conduction electrons form a DFL.

It has long been realized that generic soft modes in condensed-matter systems lead to generic scale invariance and 
nonanalyticities in the thermodynamic and transport properties as functions of the wave number, frequency, temperature, 
or external fields.\cite{Belitz_Kirkpatrick_Vojta_2005} Prominent examples include long-time-tail effects in classical 
fluids,\cite{Pomeau_Resibois_1975, Dorfman_1975, Dorfman_Kirkpatrick_Sengers_1994} the long-range nature of the 
longitudinal susceptibility in the ordered  phase of a classical Heisenberg ferromagnet,\cite{Vaks_Larkin_Pikin_1967, Brezin_Wallace_1973}
and weak-localization effects in disordered electron systems.\cite{Lee_Ramakrishnan_1985, Altshuler_Aronov_1984}
More recently, these effects have been shown to have dramatic consequences for quantum phase transitions. A prime 
example is the quantum ferromagnetic transition in (Landau) metals, which is discontinuous as a result of soft modes in a LFL 
at zero temperature, rather then continuous as the corresponding transition in classical systems.\cite{Brando_et_al_2016a}

We will focus on effects in a Dirac Metal that are universal in the sense that they are determined by the properties
of the electron system in the limit of long wavelengths and small frequencies (the ``hydrodynamic regime'') and hence
independent of detailed microscopic properties of the material other than the underlying Dirac crossing point. Of
particular importance are excitations that are soft or massless in this regime, as they can lead to nonanalytic behavior
of observables as a function of wave number, frequency, or an external field. We will discuss the existence and nature
of such soft modes in a DFL in some detail, and compare and contrast them with the analogous soft modes in a LFL.
We will then explore the consequences for the density of states, the spin susceptibility, and the specific heat in a Dirac Metal.

\section{Model}
\label{sec:II}

\subsection{Single-particle Hamiltonian, and symmetries}
\label{subsec:II.A}

A linear crossing of two bands can be described by a term $\pm v{\bm k}\cdot{\bm\sigma}$ in the single-particle Hamiltonian
in momentum space,\cite{Abrikosov_Beneslavskii_1970} where ${\bm\sigma} \equiv (\sigma_1,\sigma_2,\sigma_3)$ are the
spin Pauli matrices, and $v$ is a characteristic velocity whose value reflects the strength of the spin-orbit interaction. The
sign of this term reflects the chirality of the electrons: plus and minus describe left-handed and right-handed electrons, respectively. This term
is invariant under time reversal (TR), which changes the sign of both the momentum ${\bm k}$ and the angular momentum ${\bm\sigma}$.
Spatial inversion (SI) changes the sign of ${\bm k}$ as well as the chirality, but not the sign of ${\bm\sigma}$, so in a system with
spatial inversion symmetry the relevant part of the Hamiltonian must include both chiralities.\cite{chiralities_footnote}
If we represent the chirality degree of freedom by a second set of Pauli matrices ${\bm\pi} = (\pi_1,\pi_2,\pi_3)$, the
part of the Hamiltonian that represents the spin-orbit interaction can be written
\bse
\label{eqs:2.1}
\be
H_v =  v(\pi_3\otimes{\bm\sigma})\cdot{\bm k} \ .
\label{eq:2.1a}
\ee
We 
note that a related, but different, class of models contain a term with
formally the same structure, but with the Pauli matrices representing a pseudo-spin degree of freedom,\cite{Armitage_Mele_Vishwanath_2018}
as is the case in graphene.\cite{Kotov_et_al_2012}
The properties of the spin susceptibility are drastically different for these two classes of models, a point we will come
back to later. 

The most general single-particle Hamiltonian that is invariant under both TR and SI also includes a term that mixes the
chiralities in a symmetric way,
\be
H_{\Delta} = \Delta(\pi_1\otimes\sigma_0)\ .
\label{eq:2.1b}
\ee
with $\sigma_0$ the $2\times 2$ unit matrix in spin space, and the usual band Hamiltonian
\be
H_{\epsilon} = (\epsilon_{\bm k} - \mu)(\pi_0\otimes\sigma_0)\ .
\label{eq:2.1c}
\ee
\ese
Here 
$\pi_0$ is another copy of the $2\times 2$ unit matrix, and $\epsilon_{\bm k}$ is the single-particle energy which
 for small ${\bm k}$ is bilinear in the components of ${\bm k}$.
For simplicity we will assume $\epsilon_{\bm k}$ to be isotropic, $\epsilon_{\bm k} ={\bm k}^2/2m$, with $m$ an effective
mass. $\mu$ is the chemical potential.
$H_{\epsilon} + H_v + H_{\Delta}$ is the most general single-particle Hamiltonian for a system
that is invariant under both TR and SI and whose spin-orbit interaction is described by $H_v$; it is equivalent to the effective
Hamiltonian derived in Refs.~\onlinecite{Zhang_et_al_2009, Liu_et_al_2010} for the Bi$_2$Se$_3$ family of topological insulators. 
In our simple isotropic model we also have invariance under simultaneous rotations in spin space and real space, which
leaves ${\bm\sigma}\cdot{\bm k}$ invariant.
Coupling the fermions to an external magnetic field ${\bm h} = (0,0,h)$ via a Zeeman term (which breaks TR, of course), 
we arrive at a single-particle Hamiltonian
\begin{align}
H_0 &= (\epsilon_{\bm k} - \mu)(\pi_0\otimes\sigma_0) + v(\pi_3\otimes{\bm\sigma})\cdot{\bm k} +\Delta(\pi_1\otimes\sigma_0) 
\nonumber\\
& \hskip 10pt - h(\pi_0\otimes\sigma_3) \ .
\label{eq;2.2}
\end{align}

\subsection{Single-particle spectrum, Green function, and action for a Dirac gas}
\label{subsec:II.B}

\subsubsection{Single-particle spectrum}
\label{subsubsec:II.B.1}

The single-particle spectrum is readily obtained by finding the eigenvalues of the $4\times 4$ Hamiltonian $H$. The
single-particle energy $E_{\bm k}$ has four branches:
\bse
\label{eqs:2.3}
\bea
E_{\bm k}^{1\pm} &=& \epsilon_{\bm k} \pm\sqrt{v^2{\bm k}^2+ \Delta^2 + h^2 - 2h\sqrt{v^2 k_z^2 + \Delta^2}}\ ,
\nonumber\\
\label{eq:2.3a}\\
E_{\bm k}^{2\pm} &=& \epsilon_{\bm k} \pm\sqrt{v^2{\bm k}^2+ \Delta^2 + h^2 + 2h\sqrt{v^2 k_z^2 + \Delta^2}}\ ,
\nonumber\\
\label{eq:2.3b}
\eea
\ese
In order to discuss the various possibilities for different ranges of parameter values, let us introduce an atomic-scale
momentum $p_0$ (on the order of an inverse lattice spacing), 
velocity $v_0 = p_0/2m$, and energy $E_0 = p_0^2/2m$. We then measure $E_{\bm k}$, $\Delta$, and $h$ in
units of $E_0$, 
$v$ in units of $v_0$, and ${\bm k}$ in units of $k_0$. 
\begin{figure*}[t]
\includegraphics[width=17cm]{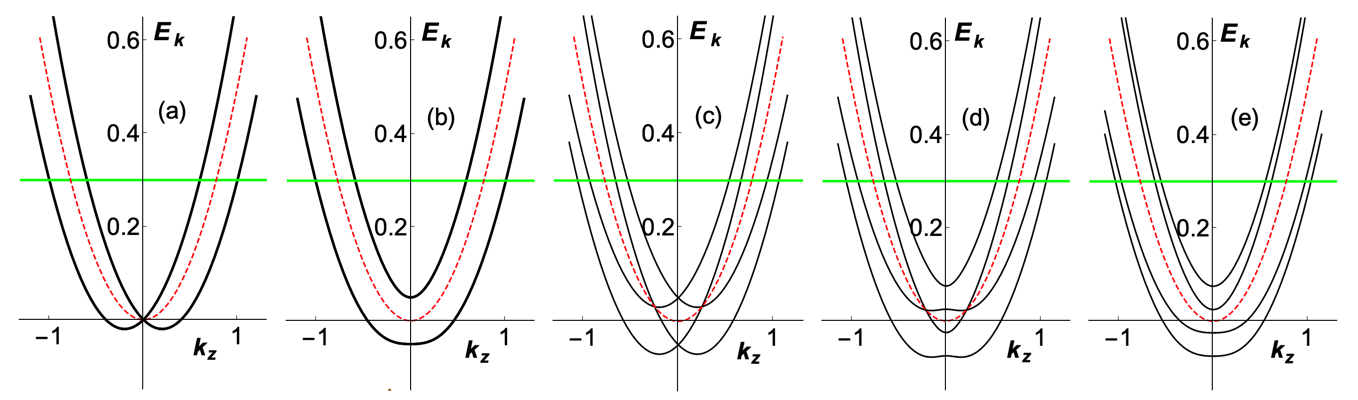}
\caption{Single-particle spectra (solid black lines) for $v=0.2$ and $\Delta=h=0$ (a), $\Delta=0.05$, $h=0$ (b), 
              $\Delta=0$, $h=0.05$ (c), $\Delta=0.025$, $h=0.05$ (d), and $\Delta=0.05$, $h=0.025$ (e), respectively. 
              The bands in panels (a) and (b) are two-fold degenerate, and the horizontal green lines denote the chemical
              potential. The parabolic spectrum for $v=\Delta=h=0$
              is shown for comparison (dashed red line). See the text for further discussion.}
\label{fig:1}
\end{figure*}
\begin{figure*}[t]
\includegraphics[width=17cm]{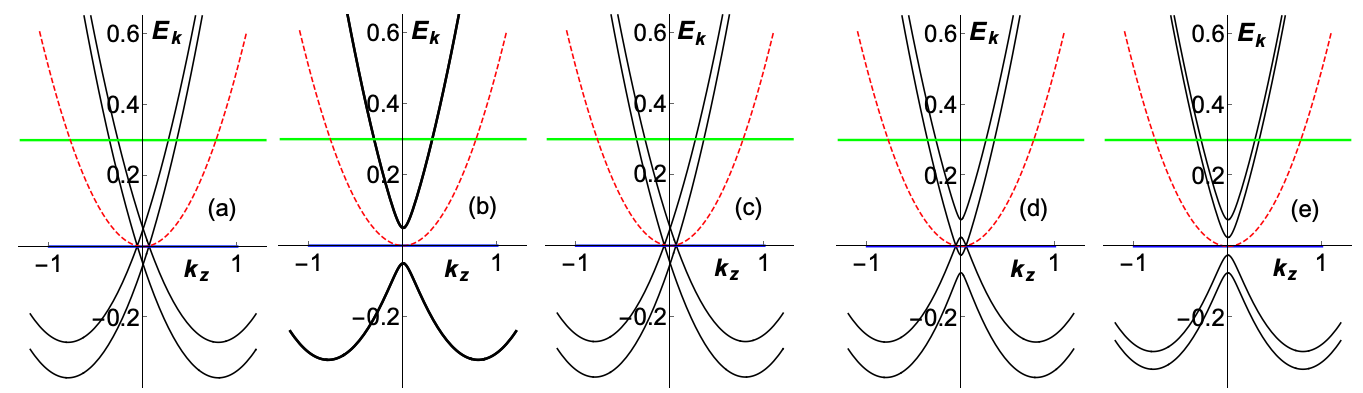}
\caption{Same as Fig.~\ref{fig:1}, but for $v=0.8$. The horizontal blue line denotes a zero chemical potential, which
              renders the system a Dirac semimetal (a), Dirac insulator (b), Weyl semimetal (c,d), and Weyl insulator (e),
              respectively. For a large chemical potential (horizontal green line) one has a Dirac (a,b) or Weyl (c-e) metal.
              See the text for a discussion of the sense in which the systems shown on Fig.~\ref{fig:1} are also propertly
              classified as Dirac
              and Weyl metals.}
\label{fig:2}
\end{figure*}

Let us start with the case of a small spin-orbit coupling. For $v=0$ the spectrum consists of four degenerate 
parabolic bands, two for each spin projection and two for each chirality. A small spin-orbit coupling $v$ splits the
bands, for $\Delta=h=0$
the Fermi surface for a nonzero chemical potential consists of two two-fold degenerate sheets, and the split
bands display a linear crossing at $k=0$, see Fig.~\ref{fig:1}(a). $\Delta\neq 0$ gaps out that band crossing,
Fig.~\ref{fig:1}(b), whereas $h\neq 0$ lifts the two-fold degeneracy and splits the crossing band pairs vertically,
Fig.~\ref{fig:1}(c).  If both $\Delta$ and $h$ are nonzero the band crossings are shifted to nonzero values of
$k$ if $h>\Delta$, and they disappear if $\Delta>h$, see Figs.~\ref{fig:1}(d,e). At the Fermi surface (for a
sizable chemical potential) all of these effects appear as a small splitting of the four-fold degenerate 
parabolic band in the absence of a spin-orbit coupling. However, as we will discuss, the properties of the 
resulting Fermi liquid are qualitatively different from the parabolic-band case and the systems can be
characterized, in a well-defined sense, as a Dirac metal (Fig.~\ref{fig:1}(a,b)) or Weyl metal 
(Fig.~\ref{fig:1}(d-e)).

For large values of $v$, when the spin-orbit term $H_v$ dominates over $\epsilon_{\bm k}$ in $H_0$, the 
spectrum can be described as follows: For $\Delta=0$ and $h=0$ it consists of two degenerate Dirac cones.
The lower cone eventually bends over, but for large $v$ this happens only for wave vectors large compared
to the inverse lattice spacing, so this branch will not contribute to the Fermi surface, see Fig.~\ref{fig:2}(a).
For a chemical potential tuned to the crossing point, $\mu=0$, the system is a Dirac semimetal, and for
$\mu>0$ it is natural to call it a Dirac metal. $\Delta>0$ gaps out the crossing point, and the system is a
Dirac insulator for $\mu<\Delta$, and a Dirac metal for $\mu>\Delta$, Fig.~\ref{fig:2}(b). $h>0$ separates
the Dirac cones in wave-number space, leading to a Weyl semimtal if $\mu=0$ and a Weyl metal if $\mu>0$,
Fig.~\ref{fig:2}(c). This structure persists for $0<\Delta<h$, Fig.~\ref{fig:2}(d), whereas for $0<h<\Delta$
the cones are gapped out and one has a Weyl insulator for $\mu=0$ and a Weyl metal for $\mu\gg\Delta$,
Fig.~\ref{fig:2}(e).

Even though the single-particle spectra in Figs.~\ref{fig:1}, \ref{fig:2} are related in obvious ways,
they appear very different at wave numbers that are not asymptotically small; this is true in particular at the Fermi
surface for generic values of the chemical potential. However, as we will show in Sec.~\ref{sec:III}, the
soft-mode spectrum is the same for both small and large values of the spin-orbit coupling $v$,\cite{large_v_footnote}
and the
systems depicted in Fig.~\ref{fig:1} therefore constitute Dirac and Weyl metals in the same sense as
those in Fig.~\ref{fig:2} for a sufficiently large value of $\mu$.  For our purposes the distinction between
the Dirac and Weyl cases is not important, and we will refer to all of the metallic systems depicted in
Figs.~\ref{fig:1}, \ref{fig:2} as `Diral metals'. We also note that in systems that are not invariant under SI,
or if TR is broken by effects other than a magnetic field, there are additional possibilities, in addition to the Dirac
and Weyl cases, that we will not discuss, see, e.g., Ref.~\onlinecite{Mitchell_Fritz_2015}.

For $\mu \gg \Delta$ and $h=0$ the various Fermi surfaces are characterized by Fermi wave numbers
that are solutions of the quadratic equations
\be
\kF^2 = 2 m^2\left(\mu/m + v^2 \pm \sqrt{(\mu/m+v^2)^2 - \mu^2/m^2}\right)\ .
\label{eq:2.4}
\ee
Our explicit calculations in Sec.~\ref{sec:IV} will be performed in the limit of a  small spin-orbit interaction, 
$v^2 \ll \mu/m$. $v$ then splits the spherical Fermi surface into two two-fold degenerate sheets, see 
Figs.~\ref{fig:1}(a,b), and we have
\bse
\label{eq:2.5}
\be
\kF^{\pm} = \sqrt{2m\mu} \pm v/2m + O(v^2) \qquad (v^2 \ll \mu/m)\ .
\label{2.5a}
\ee
A small magnetic field lifts the two-fold degeneracy, see Figs.~\ref{fig:1}(c-e). In the limit of a strong spin-orbit interaction,
$v^2 \gg \mu/m$, and for $h=0$, there is only one physical Fermi surface (see Fig.~\ref{fig:2}) with
\be
\kF = \mu/v \qquad (v^2 \gg \mu/m) \ .
\label{eq:2.5b}
\ee
\ese
The second solution of the quadratic equation, $\kF=2mv$, does not correspond to a physical
Fermi surface if it is larger than a reciprocal lattice vector.

\subsubsection{Single-particle Green function}
\label{subsubsec:II.B.2}

The single-particle Green function is defined as
\be
G_n({\bm k}) = \left[i\omega_n(\pi_0\otimes\sigma_0) - H_0\right]^{-1}\ ,
\label{eq:2.6}
\ee
with $\omega_n$ a fermionic Matsubara frequency. It is a $4\times 4$ matrix in chirality-spin space with matrix elements
$\left(G_n({\bm k})\right)_{\sigma\sigma'}^{\alpha\alpha'}$.
The complete explicit expression is quite complicated, but we will not need it for our purposes. Rather, we list various 
special cases and approximations that suffice for our objectives. 

For $\Delta = 0$ the Green function is diagonal in the chirality index and one finds
\bse
\label{eqs:2.7}
\be
G_n^{\Delta=0}({\bm k}) = \pi_+ \otimes {\mathfrak g}_n({\bm k},{\bm h}) + \pi_- \otimes {\mathfrak g}_n(-{\bm k},{\bm h})
\label{eq:2.7a}
\ee
where $\pi_{\pm} = (\pi_0 \pm \pi_3)/2$ and
\be
{\mathfrak g}_n({\bm k},{\bm h}) = \frac{(i\omega_n - \xi_{\bm k})\sigma_0 + (v{\bm k}-{\bm h})\cdot\bm\sigma}{(i\omega_n - \xi_{\bm k})^2 - (v{\bm k} - {\bm h})^2}\,
\label{eq:2.7b}
\ee
or, equivalently,
\be
\left(G_n^{\Delta=0}({\bm k})\right)^{\alpha\alpha'} =  \delta_{\alpha\alpha'}\,\frac{(i\omega_n - \xi_{\bm k})\sigma_0 + (\alpha v{\bm k}-{\bm h})\cdot{\bm\sigma}}{(i\omega_n - \xi_{\bm k})^2 - (\alpha v{\bm k} - {\bm h})^2}\ 
\label{eq:2.7c}
\ee
\ese
\begin{widetext}
Here we have defined $\xi_{\bm k} = \epsilon_{\bm k} - \mu$.  A similar simplification occurs in zero field, where one finds
\be
G_n^{h=0}({\bm k}) = \frac{1}{(i\omega_n - \xi_{\bm k})^2 - v^2{\bm k}^2 - \Delta^2} \ \bigl[(\pi_0\otimes\sigma_0)(i\omega_n - \xi_{\bm k}) + (\pi_3\otimes{\bm\sigma})\cdot v\,{\bm k} 
       + (\pi_1\otimes\sigma_0)\Delta\bigr]
\label{eq:2.8}
\ee

A nonzero field appears in both the numerator and the denominator of the Green function, as is illustrated by the special case $\Delta = 0$ in
Eqs.~(\ref{eqs:2.7}). For reasons that will become clear in Sec.~\ref{sec:IV} (see the remark before Eq.~(\ref{eq:4.7})) we do not need to retain 
the occurrences of $h$ in the numerator for our purposes. In an approximation that suffices for determining the leading nonanalytic behavior 
of the density of states and the spin susceptibility we thus keep $h$ only in the denominator and find
\be
G_n({\bm k}) \approx \frac{\left[(i\omega_n - \xi_{\bm k})^2 - (v^2{\bm k}^2 + \Delta^2)\right] \bigl[(\pi_0\otimes\sigma_0)(i\omega_n - \xi_{\bm k}) + (\pi_3\otimes{\bm\sigma})\cdot v\,{\bm k} + (\pi_1\otimes\sigma_0)\Delta\bigr]}{\left[(i\omega_n - \xi_{\bm k})^2 - (v^2{\bm k}^2 + \Delta^2 + h^2 - 2h\sqrt{v^2 k_z^2 + \Delta^2})\right]
     \left[(i\omega_n - \xi_{\bm k})^2 - (v^2{\bm k}^2 + \Delta^2 + h^2 + 2h\sqrt{v^2 k_z^2 + \Delta^2})\right]}
\label{eq:2.9}
\ee
This reduces to Eqs.~(\ref{eqs:2.7}) in the same approximation if the gap $\Delta$ is zero, or if it is is negligible compared to the additive 
terms in Eq.~(\ref{eq:2.9}). Anticipating that the wave vector ${\bm k}$ will be close to the Fermi surface, we see that the relevant criterion 
is $\Delta \ll v\kF$, with $\kF$ the Fermi wave number for the Fermi surface under consideration. This is the limit we will investigate in 
Sec.~\ref{sec:IV}. As we will discuss in Sec.~\ref{subsec:V.A} the presence of a small $\Delta$ does not qualitatively change the results.
\end{widetext}

For $\Delta=0$ the Green function is  diagonal in the chirality index, see Eq.~(\ref{eq:2.7c}). In the same approximation as in 
Eq.~(\ref{eq:2.9}), i.e., ignoring the ${\bm h}$ in the numerator, it can be written
\bse
\label{eqs:2.10}
\be
\left(G_n^{\Delta=0}\right)^{\alpha\alpha'}({\bm k}) \approx\delta_{\alpha\alpha'}\,G_n^{\alpha}({\bm k})
\label{eq:2.10a}
\ee
where
\be
G_n^{\alpha}({\bm k}) = \frac{1}{2}\left(F_n^{\alpha -}({\bm k}) M(-\alpha\hat{\bm k}) + F_n^{\alpha +}({\bm k}) M(\alpha\hat{\bm k})\right)
\label{eq:2.10b}
\ee
with
\be
F_n^{\alpha\pm}({\bm k}) = \frac{1}{i\omega_n - \xi_{\bm k} \pm \vert\alpha v {\bm k} - {\bm h}\vert}
\label{eq:2.10c}
\ee
and
\be
M(\hat{\bm k}) = \sigma_0 - {\bm\sigma}\cdot\hat{\bm k}\ .
\label{eq:2.10d}
\ee
\ese
Here $\hat{\bm k} = {\bm k}/\vert{\bm k}\vert$ is the unit vector in ${\bm k}$-direction. In writing the Green function in the form of
Eq.~(\ref{eq:2.10b}) we have performed a partial fraction decomposition of the quadratic denominator. The partial Green
function $F_n^{\alpha\pm}$ therefore has two indices: $\alpha$ is the chirality index, and the additional $\pm$ represents
the two branches of the quadratic denominator. Since spin is not a good quantum number in the presence of a spin-orbit
interaction, these two branches cannot be interpreted as representing two different spin projections.

Equations~(\ref{eqs:2.10}) no longer allow for the Landau limit of zero spin-orbit coupling to be taken, because of the approximation
involved. However, performing a partial-fraction decomposition on Eqs.~(\ref{eqs:2.7}), which are still exact, shows that the exact Landau
Green function can be written in a form analogous to Eqs.~(\ref{eqs:2.10}), viz.,
\bse
\label{eqs:2.11}
\be
\left(G_n^{v=\Delta=0}\right)^{\alpha\alpha'}({\bm k}) = \delta_{\alpha\alpha'}\,G_n({\bm k})
\label{eq:2.11a}
\ee
where
\be
G_n({\bm k}) =\frac{1}{2}\left(G_{n\downarrow}({\bm k}) M(\hat{\bm h}) + G_{n\uparrow}({\bm k}) M(-\hat{\bm h})\right)\ .
\label{eq:2.11b}
\ee
Here
\be
G_{n\uparrow\downarrow}({\bm k}) = F_n^{\alpha\pm}({\bm k})\Bigl\vert_{v=0} = \frac{1}{i\omega_n - \xi_{\bm k} \pm \vert{\bm h}\vert}\ .
\label{eq:2.11c}
\ee
\ese
are the Green functions for up-spin ($\uparrow\equiv +$) and down-spin ($\downarrow \equiv -$) electrons, respectively.

For later reference we list five properties of the matrix $M$, which can easily be checked by direct calculation:
\bse
\label{eqs:2.12}
\bea
M(\hat{\bm k}) M(\hat{\bm k}) &=& 2 M(\hat{\bm k})\ ,
\label{eq:2.12a}\\
M(\hat{\bm k}) \bm\sigma M(\hat{\bm k}) &=& -2\hat{\bm k} M(\hat{\bm k})\ ,
\label{eq:2.12b}\\
M(\hat{\bm k}) M(-\hat{\bm k}) &=& 0\ ,
\label{eq:2.12c}\\
M(\hat{\bm k}) \bm\sigma M(-\hat{\bm k}) &=& 2\left[\bm\sigma + i\hat{\bm k}\times\bm\sigma - \hat{\bm k}({\bm k}\cdot\sigma) \right]\ ,\qquad\quad
\label{eq:2.12d}\\
\tr \left(M(\hat{\bm k}) M(\hat{\bm p})\right) &=& 2(1 + \hat{\bm k}\cdot\hat{\bm p})\ .
\label{eq:2.12e}
\eea
\ese

\smallskip
\subsubsection{Action for a Dirac Fermi gas}
\label{subsubsec:II.B.3}

The action for a noninteracting Fermi system governed by the Hamiltonian $H_0$, or a Dirac Fermi gas, is given in terms of the inverse Green function,
\be
S_0 = \sum_{k} \sum_{\sigma\sigma'}\sum_{\alpha\alpha'} {\bar\psi}_{\sigma}^{\alpha}(k)\left[i\omega_n \delta_{\sigma\sigma'}\delta_{\alpha\alpha'} - (H_0)_{\sigma\sigma'}^{\alpha\alpha'}\right]
     \psi_{\sigma'}^{\alpha'}(k)\ .
\label{eq:2.13}
\ee
Here $\bar\psi$ and $\psi$ are fermionic fields with spin index $\sigma = (+,-) \equiv (\uparrow,\downarrow)$ and chirality indes $\alpha = (+,-)$,
and $k = ({\bm k},\omega_n)$ is a 4-vector that comprises the wave vector ${\bm k}$ and the Matsubara frequency $\omega_n$.
Introducing spinors $\psi^{\alpha} = (\psi_{\uparrow}^{\alpha},\psi_{\downarrow}^{\alpha})$ and a scalar product $(\bar\psi,\psi) =\sum_{\sigma} \bar\psi_{\sigma}\psi_{\sigma}$,
we can write $S_0$ more explicitly in the form
\be
S_0 = \sum_{k} \sum_{\alpha} \Bigl(\bar\psi^{\alpha}(k),\bigl((i\omega_n - \xi_{\bm k})\sigma_0 + h\sigma_3 - \alpha\,v\,{\bm\sigma}\cdot{\bm k}\bigr)\psi^{\alpha}(k)\Bigr)\ 
\nonumber\\
\ee
\vskip -20pt
\be
 + \sum_k\sum_{\alpha\neq\alpha'}  \left(\bar\psi^{\alpha}(k), \Delta\,\sigma_0\,\psi^{\alpha'}(k)\right)\ . 
 \label{eq:2.14}
\ee
To summarize the symmetry properties, this action is invariant under SI, TR (for $h=0$), and simultaneous rotations in
spin space and real space.
\begin{figure}[b]
\includegraphics[width=4.5cm]{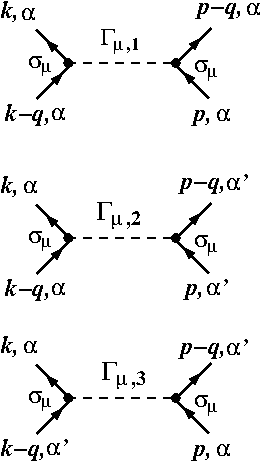}
\caption{Graphical representation of the electron-electron interaction. $\mu=0$ and $\mu=1,2,3$ represent spin-singlet
              and spin-triplet interactions, respectively. The spin-triplet amplitudes $\Gamma_{(\mu=1,2,3),i} \equiv \Gamma_{\text{t},i}$
              are identical due to spin rotational invariance. The chirality index is denoted by $\alpha$ and $\alpha'\neq\alpha$. 
              $\Gamma_{\mu,1}$ describes intra-chirality interactions, and
              $\Gamma_{\mu,2}$ and $\Gamma_{\mu,3}$ describe inter-chirality interactions.}
\label{fig:3}
\end{figure}

\subsection{Electron-electron interaction, and action for a Dirac Fermi liquid}
\label{subsec:II.C}

In order to study correlation effects, we need to add a four-fermion interaction $S_{\text{int}}$ to the noninteracting electron system
described by $S_0$. Underlying all correlations is the Coulomb interaction, but the effective interaction in a fully renormalized theory
will contain all contributions that are allowed by symmetry. In our case, the noninteracting action is invariant under simultaneous
rotations in spin space and momentum space,\cite{spin_rotational_invariance_footnote} as well as under SI; this also implies invariance 
under TR. A four-fermion interaction
that is consistent with all of these requirements can be written
\begin{widetext}
\bea
S_{\text{int}} &=& \frac{-T}{2V} \sum_{q}{}^{'}  \sum_{k,p} \Bigl[ \Gamma_{\text{s,1}} \sum_{\alpha}
     \left(\bar\psi^{\alpha}(k)\sigma_0\psi^{\alpha}(k-q)\right)\left(\bar\psi^{\alpha}(p-q)\sigma_0\psi^{\alpha}(p)\right)
\nonumber\\
   &&+\Gamma_{\text{s,2}} \sum_{\alpha\neq\alpha'}
     \left(\bar\psi^{\alpha}(k)\sigma_0\psi^{\alpha}(k-q)\right)\left(\bar\psi^{\alpha'}(p-q)\sigma_0\psi^{\alpha'}(p)\right)
   +\Gamma_{\text{s,3}} \sum_{\alpha\neq\alpha'}
     \left(\bar\psi^{\alpha}(k)\sigma_0\psi^{\alpha'}(k-q)\right)\left(\bar\psi^{\alpha'}(p-q)\sigma_0\psi^{\alpha}(p)\right)
\nonumber\\
   &&\hskip 0pt -\Gamma_{\text{t,1}} \sum_{\alpha}
     \left(\bar\psi^{\alpha}(k)\bm\sigma\psi^{\alpha}(k-q)\right)\cdot\left(\bar\psi^{\alpha}(p-q)\bm\sigma\psi^{\alpha}(p)\right)
\nonumber\\
   && -\Gamma_{\text{t,2}} \sum_{\alpha\neq\alpha'}
     \left(\bar\psi^{\alpha}(k)\bm\sigma\psi^{\alpha}(k-q)\right)\cdot\left(\bar\psi^{\alpha'}(p-q)\bm\sigma\psi^{\alpha'}(p)\right)
     -\Gamma_{\text{t,3}} \sum_{\alpha\neq\alpha'}
     \left(\bar\psi^{\alpha}(k)\bm\sigma\psi^{\alpha'}(k-q)\right)\cdot\left(\bar\psi^{\alpha'}(p-q)\bm\sigma\psi^{\alpha}(p)\right) 
             \Bigr].
             \nonumber\\
\label{eq:2.15}             
\eea
\end{widetext}
Here the prime on the sum over $q$ indicates a restriction $\vert{\bm q}\vert < \Lambda$, with $\Lambda\ll\kF$ a momentum
cutoff. This is necessary in order to avoid overcounting, as explained in Appendix~\ref{app:A}. The leading nonanalyticities
we are interested in will not depend on the cutoff.

The six terms in Eq.~(\ref{eq:2.15}) are separately invariant under the required symmetries, and come with six interaction amplitudes. They are graphically 
represented in Fig.~\ref{fig:3}. As we will see in Sec.~\ref{sec:III}, interactions that mix chiralities ($\Gamma_{\text{s},2}, 
\Gamma_{\text{s},3}, \Gamma_{\text{t},2}, \Gamma_{\text{t},3}$) play a qualitatively different role that those that do not 
($\Gamma_{\text{s},1}, \Gamma_{\text{t},1}$). A sketch of how this effective interaction is generated from a Coulomb interaction
between density fluctuations is provided in Appendix~\ref{app:A}.

The full action for our model of interacting Dirac fermions, i.e., a Dirac Fermi liquid, is given by 
\be
S_{\text{DL}} = S_0 + S_{\text{int}}\ .
\label{eq:2.16}
\ee
It still contains the Landau Fermi liquid as a limiting case if one puts the parameters $v$ and $\Delta$ in $H_0$ equal to zero.

\section{Soft Modes in a Dirac Metal}
\label{sec:III}

The nonanalytic behavior of various observables we are interested in is the result of four-fermion correlation functions that are soft or massless
at zero temperature and the dependence of these soft modes on the fields conjugate to the respective observables. Prior to performing explicit 
calculations it is helpful to discuss the existence, nature, and origin of these soft modes. We start
by discussing the soft modes in an ordinary LFL, and then generalize to the case of a DFL.
\medskip

\subsection{A Ward identity for a Landau Fermi liquid}
\label{subsec:III.A}

Consider the LFL Green function as given by Eqs.~(\ref{eqs:2.11}). It is easy to check that the following identity holds:
\bea
G_{n_1\sigma_1}({\bm k}+{\bm q}/2)\,G_{n_2\sigma_2}({\bm k}-{\bm q}/2) &=& \hskip 70pt
\nonumber\\
&& \hskip -110pt - \frac{G_{n_1\sigma_1}({\bm k}+{\bm q}/2) - G_{n_2\sigma_2}({\bm k}-{\bm q}/2)}
{i\Omega_{n_1-n_2} - {\bm k}\cdot{\bm q}/m + (\sigma_1-\sigma_2)h}
\label{eq:3.1}
\eea
where $\Omega_{n_1-n_2} = \omega_{n_1} - \omega_{n_2}$ is a bosonic Matsubara frequency, and $h = \vert{\bm h}\vert$.
This is a generalization of Velicky's Ward identity \cite{Velicky_1969} for noninteracting electrons to a nonzero magnetic field. 
Notice that the denominator on the right-had side has the structure of the differential operator of a Boltzmann equation in the
absence of the collision operator (i.e., a time derivative plus a streaming term).
It also holds, with the ballistic dynamics replaced by diffusive ones,  for electron systems with quenched disorder if one takes 
$G$ to be the unaveraged Green function and takes a disorder average on either side of 
Eq.~(\ref{eq:3.1}).\cite{Velicky_1969, Wegner_1979, diffuson_footnote} It further still holds, with a slightly more complicated
frequency structure and $m$ replaced by an appropriately renormalized effective mass, for interacting electrons at zero temperature,
with or without quenched disorder, if the left-hand side is replaced by the appropriate four-fermion correlation function that
factorizes into a product of two Green functions in the noninteracting limit.\cite{Belitz_Kirkpatrick_1997, Belitz_Kirkpatrick_2012a} 
This is a consequence of the adiabatic correspondence between single-quasiparticle states in the 
noninteracting and interacting Fermi systems, respectively, that is inherent in Landau Fermi-liquid theory. In an interacting system at nonzero temperature a
dephasing rate appears in the denominator of the right-hand side of the Ward identity and its generalization discussed
below. A nonzero temperature thus gives a mass to all of the soft modes we will discuss. However, for our purposes there is a stronger 
effect of a nonzero temperature that is most easily seen by considering convolutions of Green functions, see Eq.~(\ref{eq:3.4c}) below.

In order to see the significance of the Ward identity, analytically continue the Green function in Eq.~(\ref{eq:2.11c}) to complex frequencies
$z$ and consider its analytic structure. The function $G_{\pm}({\bm k},z) = 1/(z - \xi_{\bm k} \pm \vert{\bm h}\vert)$ has a cut on 
the real axis, and close to the real axis it is given by 
\be
G_{\pm}({\bm k},\omega\pm i0) = \frac{1}{\omega - \xi_{\bm k} \pm \vert h\vert} \mp i\pi \delta(\omega - \xi_{\bm k} \pm \vert{\bm h}\vert)
\label{eq:3.2}
\ee
We see that the behavior of the right-hand side of Eq.~(\ref{eq:3.1}) is drastically different depending on whether the two fermionic
Matsubara frequencies $\omega_{n_1}$ and $\omega_{n_2}$ have the same sign, or opposite signs. In latter case, the numerator
on the right-hand side is nonzero in the limit $\Omega_{n_1-n_2}\to 0$, ${\bm q}\to 0$ for all frequencies or energies where the
spectrum of the Green function is nonzero. The denominator, on the other hand, is small of $O(\epsilon)$, where $\epsilon$
stands for $\Omega$, or $q = \vert{\bm q}\vert$, or $h$, all of which scale the same way. 
In particular, it vanishes in the limit of zero frequency $\Omega$ and wave number $q$ if the two spin projections are the same
($\sigma_1 = \sigma_2$). The four-fermion correlation on the left-hand-side is thus a soft mode of $O(1/\epsilon)$. Whereas, if $n_1$ and
$n_2$ have the same sign, then both the numerator and the denominator are small of $O(\epsilon)$, and the four-fermion
correlation is of $O(1)$. Similarly, products of $N$ Green functions $G_{n_1} G_{n_2} \ldots G_{n_N}$ are of $O(1/\epsilon^{N-1})$
if the $N$ frequency indices $n_i$ do not all have the same sign, but of $O(1)$ if they do.

It is illustrative to derive the same result as a Ward identity that results from a unitary transformation of the fermionic
fields that represents a rotation in frequency space,\cite{Wegner_1979} see Appendix~\ref{app:B}. This shows that the
soft four-fermion excitations can be interpreted as the Goldstone modes of a spontaneously broken continuous symmetry,
viz., the symmetry between retarded (positive Matsubara frequency) and advanced (negative Matsubara frequency) degrees
of freedom. The relevant order parameter is the spectrum of the Green function. 

\bigskip
\subsection{A Ward identity for a Dirac Fermi liquid}
\label{subsec:III.B}

We now consider the analogous identity involving the Green function $F_n^{\alpha\beta}({\bm k})$ given in Eq.~(\ref{eq:2.10c}).
We find
\begin{widetext}

\be
F_{n_1}^{\alpha_1 \beta_1}({\bm k}+{\bm q}/2) F_{n_2}^{\alpha_2\beta_2}({\bm k}-{\bm q}/2) = - \frac{F_{n_1}^{\alpha_1 \beta_1}({\bm k}+{\bm q}/2) - F_{n_2}^{\alpha_2\beta_2}({\bm k}-{\bm q}/2)}
   {i\Omega_{n_1-n_2} - {\bm k}\cdot{\bm q}/m + \beta_1\vert\alpha_1 v ({\bm k}+{\bm q}/2) - {\bm h}\vert - \beta_2\vert\alpha_2 v ({\bm k}-{\bm q}/2) - {\bm h}\vert  }
\label{eq:3.3}
\ee
For the numerator, the same discussion as in the LFL case applies, but the denominator now allows for more possibilities.
Anticipating that in perturbation theory with respect to the electron-electron interaction we will encounter convolutions of Green functions 
we define a correlation function
\bse
\label{eqs:3.4}
\be
\varphi^{\alpha_1\alpha_2}_{\beta_1\beta_2}({\bm q},i\Omega_n;i\omega_m) = \frac{1}{V}\sum_{\bm k} F_k^{\alpha_1\beta_1} F_{k+q}^{\alpha_2\beta_2}
\label{eq:3.4a}
\ee
where $k=({\bm k},i\omega_m)$, $q=({\bm q},i\Omega_n)$, and $F_k^{\alpha\beta} = F_n^{\alpha\beta}({\bm k})$. Rather than performing
the integral exactly, we note that the singular structure at small momenta and wave numbers we are interested in is preserved in the
well-known approximation that performs the radial part of the momentum integral by means of a contour integration over the
interval $-\infty < \xi_{\bm k} < \infty$.\cite{Abrikosov_Gorkov_Dzyaloshinski_1963} Within this scheme, which we refer to as
the AGD approximation, we are left with only an angular integral,
\be
\varphi^{\alpha_1\alpha_2}_{\beta_1\beta_2}({\bm q},i\Omega_n; i\omega_m) = \NF \int \frac{d\Omega_{\bm k}}{4\pi} \frac{-2\pi i \, \sgn(\omega_m) \, \Theta(-\omega_m(\omega_m + \Omega_n))}{i\Omega_n - \vF{\hat{\bm k}}\cdot{\bm q}
   + \beta_2\vert\alpha_2 v\kF \hat{\bm k} + \alpha_2 v {\bm q} - {\bm h}\vert - \beta_1\vert\alpha_1 v\kF \hat{\bm k} - {\bm h}\vert}\ .
\label{eq:3.4b}
\ee
Here $\Omega_{\bm k}$ is the solid angle with respect to the wave vector ${\bm k}$ (not to be confused with the bosonic Matsubara
frequency $\Omega_n$), $\vF = \kF/m$ and $\NF = \kF m/2\pi^2$ are the Fermi velocity and the density of states per spin and chirality, 
respectively, at the Fermi surface under consideration,\cite{Fermi_surface_footnote} and the expression is valid for the leading singular behavior only.

Depending on the relative values of $\alpha_{1,2}$ and $\beta_{1,2}$ there may or may not be a mass in the denominator of Eq.~(\ref{eq:3.4b}).
Before we discuss the various cases, we consider the dependence on the Matsubara frequency $\Omega_n$ and its consequences for the
soft-mode structure at nonzero temperature. To this end it is useful to perform the sum over the fermionic Matsubara frequency. We define
\be
\varphi^{\alpha_1\alpha_2}_{\beta_1\beta_2}({\bm q},i\Omega) = T\sum_{i\omega_m} 
  \varphi^{\alpha_1\alpha_2}_{\beta_1\beta_2}({\bm q},i\Omega_n; i\omega_m)
     =  \NF \int \frac{d\Omega_{\bm k}}{4\pi} \frac{i\Omega_n}{i\Omega_n - \vF{\hat{\bm k}}\cdot{\bm q}
   + \beta_2\vert\alpha_2 v\kF \hat{\bm k} + \alpha_2 v {\bm q} - {\bm h}\vert - \beta_1\vert\alpha_1 v\kF \hat{\bm k} - {\bm h}\vert}\ .
\label{eq:3.4c}
\ee
\ese
\end{widetext}
%
The factor of $i\Omega_n$ in the numerator is a consequence of the frequency-mixing requirement, which in Eq.~(\ref{eq:3.4b}) was
reflected by the $\Theta$ function: Only products of retarded and advanced Green functions can produce convolutions that are potentially soft,
which makes the frequency summation trivial. In any frequency sum over products
of convolutions the term with $n=0$ therefore comes with zero weight, and the smallest frequency index that contributes is $n=1$. The $i\Omega_n$
in the denominator at nonzero temperature thus effectively acts as a mass that is linear in $T$. This is a stronger effect than the dephasing rate
mentioned after Eq.~(\ref{eq:3.1}), which vanishes faster than $T$ for $T\to 0$, and it is responsible for the temperature scaling as the wave
number and the magnetic field.

\subsection{Soft modes in Landau and Dirac Fermi liquids}
\label{subsec:III.C}

We now discuss two limiting cases.

{\it First case: $h\gg v\kF$.} To zeroth approximation we can put $v=0$ and recover the LFL case. $\varphi$ is then independent of the
chirality index, and $\beta = \pm \equiv \uparrow\downarrow$ labels the spin projection. Consistent with Eq.~(\ref{eq:3.1}), we have
two types of singular behavior. The first type is realized by the $\uparrow\downarrow$ and $\downarrow\uparrow$ combinations:
\bse
\label{eqs:3.5}
\be
\varphi_{\beta,-\beta} ({\bm q},i\Omega_n) = \NF \int \frac{d\Omega_{\bm k}}{4\pi} \frac{i\Omega_n}{i\Omega_n - \vF{\hat{\bm k}}\cdot{\bm q} -2\beta h }\ .
\label{eq:3.5a}
\ee
It is characterized by singular behavior for $\Omega, \vert{\bm q}\vert \to 0$ that is cut off by a magnetic field $h$ (which scales as $\Omega$ and $\vert{\bm q}\vert$).
We will refer to soft modes of this type as ``soft modes of the first kind" (with respect to $h$, see Sec.~\ref{subsec:III.D} below).
The second kind is realized by the $\uparrow\uparrow$ and $\downarrow\downarrow$ combinations,
\be
\varphi_{\beta\beta}({\bm q},i\Omega_n) = \NF \int \frac{d\Omega_{\bm k}}{4\pi} \frac{ i\Omega_n}{i\Omega_n - \vF{\hat{\bm k}}\cdot{\bm q} }\ ,
\label{eq:3.5b}
\ee
\ese
which we recognize as the hydrodynamic part of the Lindhard function. Here the hydrodynamic singularity is {\em not} cut off by a magnetic field. 
We will refer to this type as ``soft modes of the second kind".  


{\it Second case: $h \ll v\kF$.} In this case we can ignore $h$ compared to $v\kF$ in Eq.~(\ref{eq:3.4b}). The correlations with $\beta_1 = -\beta_2$
then become independent of the chirality index, and we have
\bse
\label{eqs:3.6}
\be
\varphi_{\beta,-\beta}^{\alpha_1\alpha_2} ({\bm q},i\Omega_n) = i\Omega_n\,\NF \int \frac{d\Omega_{\bm k}}{4\pi} \frac{1}{i\Omega_n - \vF{\hat{\bm k}}\cdot{\bm q} -2\beta v\kF }\ .
\label{eq:3.6a}
\ee
The spin-orbit interaction thus gives a mass to the modes that were soft of the first kind in the LFL case,
see Eq.~(\ref{eq:3.5a}). The correlations with $\beta_1 = \beta_2$, on the other hand, remain soft:
\bea
\varphi_{\beta\beta}^{\alpha_1\alpha_2} ({\bm q},i\Omega_n) &=& i\Omega_n\,\NF \int \frac{d\Omega_{\bm k}}{4\pi} 
\nonumber\\
&& \hskip -60pt \times \frac{1}{i\Omega_n - (\vF - \beta v){\hat{\bm k}}\cdot{\bm q} 
     + \beta(\alpha_1 - \alpha_2){\hat{\bm k}}\cdot{\bm h} }\ .\qquad
\label{eq:3.6b}
\eea
\ese
They are soft of the first kind for correlations that mix the chiralities ($\alpha_1 \neq \alpha_2$), and soft of the second kind
for correlations within a given chirality ($\alpha_1 = \alpha_2$). We note that the soft modes of the first kind have a chirality
structure that is different from the one necessary for forming a density of a spin density, which is diagonal in $\alpha$:
\be
{\bm n}_{\text{s}}(q) =  \sum_{k,\alpha}  \Bigl(\bar\psi^{\alpha}(k),\bm\sigma \psi^{\alpha}(k-q)\Bigr)
\label{eq:3.7}
\ee
These are nevertheless the modes that produce a nonanalyticity in the spin susceptibility, as we will discuss
next.

We also note that models of the graphene-type mentioned after Eq.~(\ref{eq:2.1a}), while allowing for a formally
very similar discussion, lead to a drastically different physical interpretation. If the Pauli matrix in the $\bm\sigma\cdot{\bm k}$
term is a pseudo-spin, then the coupling constant $v$ does not give a mass to the modes that were soft of the first kind
in a LFL. This has important consequences for the effects of the soft modes on the spin susceptibility, as we will discuss
in Sec.~\ref{sec:V}.

\subsection{Physical consequences of soft modes: Nonanalytic behavior of observables}
\label{subsec:III.D}

In order to determine the effects of the soft modes on observables, we need to generalize the notion of soft modes of the first
and second kinds from the last subsection. Consider an observable $\cal O$ and its conjugate field $\mathfrak h$. Each soft
mode can then be classified as being of the the first or second kind with respect to $\mathfrak h$, depending on whether or
not $\mathfrak h$ cuts off the singularity. In Sec.~\ref{subsec:III.C}, $\mathfrak h$ was the magnetic field $h$, and $\cal O$ was
the spin density. As illustrated by Eq.~(\ref{eq:3.6b}), the correlations we denoted by $\varphi_{\beta\beta}^{\alpha\alpha'}$ are
of the first kind with respect to $h$ if $\alpha \neq \alpha'$, and of the second kind if $\alpha = \alpha'$. However, if we take
$\mathfrak h$ to be the chemical potential $\mu$, which makes $\cal O$ the number density, then all of the soft modes are
of the second kind with respect to $\mu$. Now consider the partition function $Z$ as a generating functional $Z[\mathfrak h]$
that depends on $\mathfrak h$. Then the free energy is given by $f[\mathfrak h] = -(T/V)\ln Z[\mathfrak h]$, the observable by
${\cal O} = \partial f/\partial{\mathfrak h}$, and the corresponding susceptibility by $\chi_{\cal O} = \partial^2 f/\partial{\mathfrak h}^2$. 
Now $f$, and all of its derivatives, will be nonanalytic functions of $\mathfrak h$ if some mode is soft of the first kind with
respect to $\mathfrak h$. Generically, the field will scale linearly with the frequency or temperature, and with the external
wave number, and scaling implies that $\chi_{\cal O}$ will also be a nonanalytic function of those. 
Necessary conditions for these nonanalyticities to be realized are of course the mixing of positive and negative frequencies, and 
a coupling to the relevant soft modes of the first kind. The first condition requires an interacting fermion system, and the second
one implies that only interaction amplitudes in certain channels will lead to nonanalyticities.

We will now apply these considerations, which were first discussed in Ref.~\onlinecite{Belitz_Kirkpatrick_Vojta_2002},
to various specific observables. We will first show that they correctly reproduce the known nonanalyticities in a LFL, and
then use them to predict the corresponding results in a DFL.

\subsubsection{Nonanalyticities in a Landau Fermi liquid revisited}
\label{subsubsec:III.D.1}

Consider the static spin susceptibility $\chi_s = \partial^2 f/\partial h^2$ in a clean LFL. In the absence of an interaction there is no frequency mixing, 
and hence $\chi_s$ in a Fermi gas, i.e., the Lindhard function, is analytic at zero wave number and zero field. It is easy to see in
perturbation theory that there still is no frequency mixing to first order in the interaction amplitudes, but starting at second order
an electron-electron interaction mixes retarded and advanced degrees of freedom, and the relevant soft modes of the first kind are given by 
Eq.~(\ref{eq:3.5a}). They involve a mixing of the spin projections, and therefore at least one spin-triplet interaction amplitude $\Gamma_t$ 
(more precisely, one of the two components of the spin triplet that are transverse with respect to the field) is required in order to produce
a nonanalytic behavior of $\chi_s$. This can be seen from Eq.~(\ref{eq:2.15}): Only the terms that involve the Pauli matrices
$\sigma^1$ and $\sigma^2$ have the requisite $\uparrow\downarrow$ spin structure. We thus expect $\chi_s$ in a LFL to be a 
nonanalytic function of the wave number $Q$, the magnetic field $h$, and also the temperature (which scales the same as the frequency) with a 
prefactor of the nonanalyticity that for weak interactions is quadratic in the $\Gamma$ and contains at least one $\Gamma_t$. 
Simple scaling arguments\cite{Belitz_Kirkpatrick_2014} show that the leading nonanalyticity takes the form
$\delta\chi_s = \chi_s(Q) - \chi_s(Q=0) \propto Q^{d-1}$ in generic spatial dimensions $d$, and $\delta\chi_s \propto Q^2\ln Q$ in $d=3$.
Using only the soft-mode structure of the LFL we thus conclude that the spin susceptibility as a function of either $Q$, or $h$, of $T$,
with the other two variables equal to zero, has a nonanalytic behavior
\be
\delta\chi_s \propto \begin{cases} Q^{d-1}\quad\text{for}\quad h=T=0\\
                                                      h^{d-1}\quad\text{for}\quad Q=T=0\\
                                                      T^{d-1}\quad\text{for}\quad Q=h=0\ ,
                               \end{cases}
\label{eq:3.8}
\ee  
in generic dimensions, and $\delta\chi_s \propto Q^2\ln Q$ etc. in $d=3$.                            

This is indeed the case, as is well known from explicit calculations,\cite{Belitz_Kirkpatrick_Vojta_1997,
Betouras_Efremov_Chubukov_2005} with one exception: In 3-d systems the leading temperature nonanalyticity at $O(\Gamma_t^2)$
has a zero prefactor.\cite{Carneiro_Pethick_1977} This zero is accidental, however, and has no bearing on the general argument. Analogous considerations
and results hold for LFLs in the presence of quenched disorder,\cite{Altshuler_Aronov_1984, Belitz_Kirkpatrick_1994} only the
soft modes are diffusive rather than ballistic. As a result, the singularity is stronger, $\delta\chi_s \propto Q^{d-2}$ etc, and the
prefactors are linear in $\Gamma_t$ for weak interactions.

It is also illustrative to note that none of the soft modes are cut off by the chemical potential, which is the field conjugate to the
electron number density. The density susceptibility $\chi_n = \partial n/\partial\mu = \partial^2 f/\partial\mu^2$ therefore can 
{\em not} be a nonanalytic function of the wave number or the temperature. The same is true for the susceptibilities of the density
current, as well as the related higher-rank tensor correlation functions. For instance, a source term $({\bm J}\cdot{\bm k})\,\bar\psi(k)\psi(k)$
in the action generates the number-current density. Such a term can be absorbed into the single-particle energy $\epsilon_{\bm k}$
by completing the square, and hence does not cut off any of the soft modes. This is also confirmed by explicit 
calculations,\cite{Belitz_Kirkpatrick_Vojta_1997} and we will discuss the consequences in Sec.~\ref{sec:V}.

The extreme opposite case is represented by a nonzero temperature, which cuts off {\em all} of the soft modes, see the
discussion after Eq.~(\ref{eq:3.4b}) above. 
Focusing on $d=3$ from here on, we thus expect the
specific-heat coefficient $\gamma_V = \partial^2 f/\partial T^2$ in a clean 3-d LFL 
to have a $T^2\ln T$ nonanalyticity  to which both spin-singlet and spin-triplet soft modes contribute. 
This is indeed the case,\cite{Chubukov_Maslov_Millis_2006} and an analogous result holds for disordered electrons.\cite{Belitz_Kirkpatrick_1994}
In a magnetic field two of the three spin-triplet modes become massive, and the corresponding $T^2\ln T$ contributions
turn into $h^2\ln h$ terms.

We finally consider the density of states, which is given in terms of the spectrum of the Green function. It is not a thermodynamic quantity but can be 
generated if we add to the action a source term that has the structure of a frequency-dependent chemical potential:
\bea
S_{\delta\mu} &=& \sum_{n} \delta\mu(\omega_n)\frac{1}{V}\sum_{\bm k} \sum_{\sigma} \sum_{\alpha} \left(\bar\psi_{\sigma}^{\alpha}({\bm k},\omega_n) \psi_{\sigma}^{\alpha}({\bm k},\omega_n)\right.
\nonumber\\
&& - \left. (\bar\psi_{\sigma}^{\alpha}({\bm k},-\omega_n) \psi_{\sigma}^{\alpha}({\bm k},-\omega_n)\right)
\label{eq:3.9}
\eea
with $\delta\mu(\omega_n) = -\delta\mu(-\omega_n)$ a source field that is an odd function of the Matsubara frequency and discontinuous at $\omega_n=0$,
$\delta\mu(0+) = -\delta\mu(0-) \neq 0$. It is not a physically realizable field, but generates the density of states $N$ at the Fermi level via
\be
N = \frac{i}{2\pi}\,\lim_{\omega_n\to 0+} \ \frac{\partial}{\partial\, \delta\mu(\omega_n)}\Bigr\vert_{\delta\mu=0} \ln Z[\delta\mu]
\label{eq:3.10}
\ee
This source field gives all of the soft modes a mass, just as a nonzero temperature does: In the denominator on the right-hand side of the Ward identity, 
Eq.~(\ref{eq:3.1}) or (\ref{eq:3.3}), a term $\delta\mu(\omega_{n_1}) - \delta\mu(\omega_{n_2})$ appears, which for $n_1\to 0+$, $n_2\to 0-$
becomes $2\delta\mu(0+)$.

We thus expect the density of states, as a function of the frequency
or energy $\omega$ at zero temperature, to have a leading nonanalytic contribution that is proportional to $\omega^2\ln\omega$, or
to $T^2\ln T$ for the density of states on the Fermi surface as a function of the temperature, and with both the spin-singlet and the spin-triplet
modes contributing. As in the case of the specific-heat coefficient, a magnetic field will give two of the three spin-triplet modes a mass and
turn the corresponding $T^2\ln T$ or $\omega^2\ln \omega$ into $h^2\ln h$.
In generic dimensions $1<d<3$, all of these nonanalyticities turn into power laws with an exponent $d-1$.

\subsubsection{Nonanalyticities in a Dirac Fermi liquid}
\label{subsubsec:III.D.2}

In the preceding subsection we saw that one can deduce the known results for nonanalyticities in a LFL from the structure of the soft modes
alone. We will now employ the same line of reasoning to deduce the nature of the nonanalyticities in a DFL. 
For simplicity, we will restrict ourselves to $d=3$. It is obvious from power counting what the corresponding results are
for $1<d<3$.
In Sec.~\ref{sec:IV} we will 
demonstrate that explicit calculations of the spin and density susceptibilities, as well as the density of states, confirm these results.

For the density susceptibility we obtain the same null result as in the LFL case: The chemical potential $\mu$, which is the field conjugate to the
particle number, gives none of the soft modes a mass. The free energy therefore is an analytic function of $\mu$, and the density susceptibility
$\partial n/\partial\mu$ has no nonanalyticities.

The fields conjugate to the density of states and the specific-heat coefficient, viz., the frequency-dependent chemical potential and the temperature, 
respectively, give all of the soft modes a mass. The density of states will thus have a nonanalytic contribution
\bse
\label{eqs:3.11}
\bea
\delta N(\omega,T=0,h=0) &\propto& \NF (\omega/\epsilonF)^2 \ln(\vert\omega\vert/\epsilonF)\qquad
\label{eq:3.11a}\\
\delta N(\omega=0,T,h=0) &\propto& \NF (T/\epsilonF)^2 \ln(T/\epsilonF)
\label{eq:3.11b}
\eea
where $\omega$ is the energy measured from the Fermi surface. All of the coupling constants defined in Sec.~\ref{subsec:II.C} will contribute
to the prefactors. A magnetic field gives modes that couple different chiralities a mass, see Eq.~(\ref{eq:3.6b}), and we expect for the
corresponding contributions
\be
\delta N(\omega=0,T=0,h) \propto \NF (h/\epsilonF)^2 \ln(h/\epsilonF)
\label{eq:3.11c}
\ee
\ese
Contributions that do not mix chiralities will remain unaffected by the field. By the same argument, we obtain for the
specific-heat coefficient
\bse
\label{eqs:3.12}
\bea
\delta\gamma_V(T,h=0) &\propto& \NF (T/\epsilonF)^2 \ln(T/\epsilonF)
\label{eq:3.12a}\\
\delta\gamma_V(T=0,h) &\propto& \NF (h/\epsilonF)^2 \ln(h/\epsilonF)
\label{eq:3.12b}
\eea
\ese
with all interaction amplitudes contributing to the former result, but only the $\Gamma_{\mu,(2,3)}$ to the latter.

For the spin susceptibility the relevant soft modes are the ones that are made massive by a nonzero magnetic field.
From Eq.~(\ref{eq:3.6b}) we see that the only soft modes that contribute are those that mix chiralities. We thus
expect
\bse
\label{eqs:3.13}
\bea
\chi_s(Q,T=0,h=0) &\propto& \NF\,(Q/\kF)^2 \ln(q/\kF)\qquad
\label{eq:3.13a}\\
\chi_s(Q=0,T,h=0) &\propto& \NF\,(T/\epsilonF)^2 \ln(T/\epsilonF)
\label{eq:3.13b}\\
\chi_s(Q=0,T=0,h) &\propto& \NF\,(h/\epsilonF)^2 \ln(h/\epsilonF)
\label{eq:3.13c}
\eea
\ese
In all three cases the prefactors will depend only on the inter-chirality interaction amplitudes. Note that the
modes that caused the nonanalyticities in $\chi_s$ in a LFL are massive in a DFL, see Eq.~(\ref{eq:3.6a}). As a result, an
interaction that couples different chiralities is necessary in order to obtain any nonanalytic behavior of $\chi_s$ in a DFL.

\section{Nonanalytic Behavior of Observables: Explicit Calculations}
\label{sec:IV}

We now demonstrate that the results predicted by the general arguments in Sec.~\ref{subsubsec:III.D.2} are
confirmed by explicit perturbative calculations. 
We perform the calculations for the case of a spin-orbit interaction that is weak in the sense $v\ll\sqrt{\mu/m}$.
In particular, we ignore the
differences between the various values of $\kF$, $\vF$, and $\NF$. We will ignore all contributions from
modes that are massive due to $v>0$. This implies that our results are observable for wave numbers $q/\kF \alt v/\vF$,
see Eq.~(\ref{eq:3.6a}). In particular, this implies that we will not be able to recover the LFL limit by taking $v\to 0$.
We will further work for $\Delta=0$; as we will show in Sec.~\ref{subsec:V.A} the results remain qualitatively valid for 
$\Delta\neq 0$ as long as $\Delta \ll v\kF$.\cite{Zero-gap_footnote} We start with the spin susceptibility of a DFL. 

\subsection{Spin susceptibility}
\label{subsec:IV.A}

\begin{figure}[t]
\includegraphics[width=8cm]{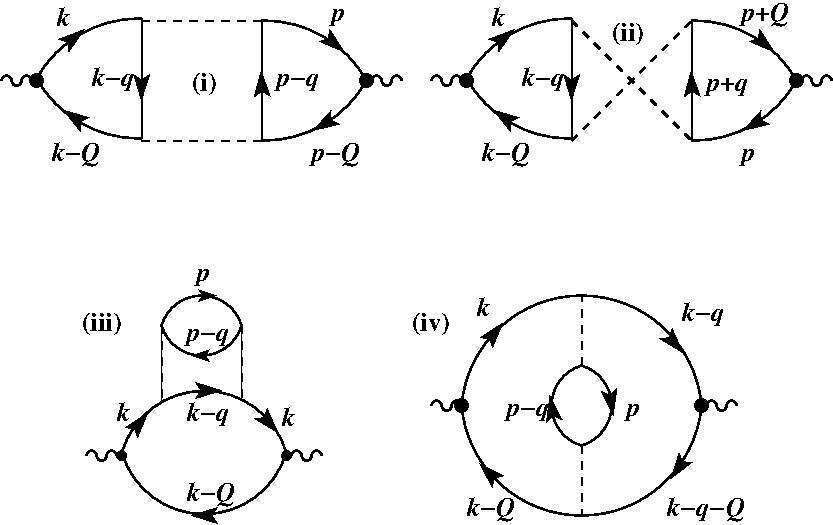}
\caption{Contributions to the spin susceptibility at second order in the interaction amplitudes. Directed solid lines represent
              Green functions, dashed lines represent interaction amplitudes. The external vertex (heavy dot) represents a Pauli 
              matrix $\sigma_3$. The internal vertices carry Pauli matrices that depend on the spin channel represented by the
              corresponding dashed line. Diagram (iii) carries a multiplication factor of $2$.}
\label{fig:4}
\end{figure}

The spin susceptibility is given by the two-point correlation function of the spin density, Eq.~(\ref{eq:3.7}). 
We consider the static longitudinal susceptibility
\be
\chisL(Q) = \langle n_{\text{s}}^3(Q)\, n_{\text{s}}^3(-Q)\rangle
\label{eq:4.1}
\ee
with $Q = ({\bm Q},i0)$ the external wave vector/frequency. We
are interested in the dependence of $\chisL$ on the external wave number $ \vert{\bm Q}\vert$, the magnetic field $h$,
and the temperature $T$. Any nonanalytic dependence on these parameters requires that one goes to at least second order in
perturbation theory with respect to the electron-electron interaction, and we denote this contribution to the spin susceptibility by
$\delta\chisL$. With the interaction given by Eq.~(\ref{eq:2.15}), and to lowest order in a loop expansion, there are four structurally 
different diagrams that contribute to $\delta\chisL$, which are shown in Fig.~\ref{fig:4}.\cite{two-loop_footnote}
%
As mentioned above, we keep only contributions that are not made
massive by the spin-orbit coupling $v$ via Eq.~(\ref{eq:3.6a}). That is, we keep only Green function convolutions where all of 
the Green functions $F^{\alpha\beta}$ share the same branch index $\beta$.

\subsubsection{Wave-number dependence}
\label{subsubsec:IV.A.1}

We first consider the dependence on the external wave number $Q$ for $h=T=0$. Diagrams (i) and (ii) in Fig.~\ref{fig:4}
lead to products of convolutions of three Green functions each. They can be realized with the interaction amplitudes
$\Gamma_{\text{s},1}$ and $\Gamma_{\text{t},1}$, or with $\Gamma_{\text{s},2}$ and $\Gamma_{\text{t},2}$ (the
$\Gamma_{\mu,1}$ and $\Gamma_{\mu,2}$ do not mix). From the discussion in Sec.~\ref{subsec:III.D} we do not expect
these processes to contribute, since the corresponding soft modes are of the first kind with respect to $h$. Indeed,
the contributions from the individual diagrams cancel due to the symmetry properties of convolutions of Green
functions listed in Appendix~\ref{app:C}: Diagrams (i) and (ii), if realized with $\Gamma_{\mu,1}$ and $\Gamma_{\mu,2}$,
cancel due to Eq.~(\ref{eq:C.1b}), and diagrams (iii) and (iv) cancel due to Eq.~(\ref{eq:C.3c}).

This leaves $\Gamma_{\mu,3}$ as the only candidates for producing singular behavior, as expected from the general
arguments in Sec.~\ref{subsec:III.D}. An analysis of the spin structure of the diagrams alone shows that any realization
that contains a $\Gamma_{\text{s},3}$ cannot contribute to the leading singularities since the leading terms contain
spin-matrix products of the type shown in Eq.~(\ref{eq:2.12c}), which vanish. Therefore, the only terms that
require a full calculation are ones proportional to $(\Gamma_{\text{t},3})^2$.

Realizing the diagrams in Fig.~\ref{fig:4} by means of $\Gamma_{\text{t},3}$ leads to the following expression for
$\delta\chisL$ in terms of integrals:
\begin{widetext}
\bse
\label{eqs:4.2}
\bea
\delta\chisL({\bm Q}) &=& \frac{1}{4}\,(\Gamma_{t,3})^2 \sum_{q}{}^{'}\sum_{k,p} 
     \biggl\{ {\hat k}_z {\hat p}_z \left[1+(\bm{\hat k}\cdot\bm{\hat p})^2\right] \sum_{\beta}\beta F_k^{+\beta} F_{k-q}^{-\beta} F_{k-Q}^{+\beta}
     \sum_{\beta'}\beta' \left(F_p^{+\beta'} F_{p-q}^{-\beta'} F_{p-Q}^{+\beta'} - F_p^{-\beta'} F_{p+q}^{+\beta'} F_{p+Q}^{-\beta'}\right)
\nonumber\\
&& \hskip 103pt + 2 {\hat k}_z {\hat p}_z (\hat{\bm k}\cdot\hat{\bm p}) \sum_{\beta} F_k^{+\beta} F_{k-q}^{-\beta} F_{k-Q}^{+\beta}
     \sum_{\beta'} \left(F_p^{+\beta'} F_{p-q}^{-\beta'} F_{p-Q}^{+\beta'} - F_p^{-\beta'} F_{p+q}^{+\beta'} F_{p+Q}^{-\beta'}\right)
\nonumber\\
&& \hskip 40pt + {\hat k}_z^2 \left[1+(\bm{\hat k}\cdot\bm{\hat p})^2\right] 
       \sum_{\beta} \left( 2 F_k^{+\beta} F_k^{+\beta} F_{k-q}^{-\beta} F_{k-Q}^{+\beta} - F_k^{+\beta} F_{k-q}^{-\beta} F_{k-Q}^{+\beta} F_{k-q-Q}^{-\beta} \right)          
       \sum_{\beta'} F_p^{+\beta'} F_{p-q}^{-\beta'}
\nonumber\\
&& \hskip 65pt +2 {\hat k}_z^2 (\hat{\bm k}\cdot\hat{\bm p}) \sum_{\beta} \beta 
                    \left( 2 F_k^{+\beta} F_k^{+\beta} F_{k-q}^{-\beta} F_{k-Q}^{+\beta} - F_k^{+\beta} F_{k-q}^{-\beta} F_{k-Q}^{+\beta} F_{k-q-Q}^{-\beta}\right)
                           \sum_{\beta'} \beta' F_p^{+\beta'} F_{p-q}^{-\beta'} \biggr\}
\nonumber\\
&& \hskip 125pt +\ (\text{the same terms with ${\bm Q}\to -{\bm Q}$})
\label{eq:4.2a}
\eea
where $F_k^{\alpha\beta} \equiv F_n^{\alpha\beta}({\bm k})$ is given by Eq.~(\ref{eq:2.10c}).
This expression is valid only for the leading singular contributions to $\delta\chisL$. The first two terms come from diagrams (i) and (ii), 
and the last two from (iii) and (iv). By using the symmetry relations expressed in Eqs.~(\ref{eq:C.2b}) and (\ref{eq:C.4c}) we can
rewrite this result as follows:
\bea
\delta\chisL({\bm Q}) &=& \frac{1}{2}\,(\Gamma_{t,3})^2 \sum_{q}{}^{'}\sum_{k,p} 
     \biggl\{ {\hat k}_z {\hat p}_z \left[1+(\bm{\hat k}\cdot\bm{\hat p})^2\right] \sum_{\beta}\beta F_k^{+\beta} F_{k-q}^{-\beta} F_{k-Q}^{+\beta}
     \sum_{\beta'}\beta' F_p^{+\beta'} F_{p-q}^{-\beta'} F_{p-Q}^{+\beta'} 
\nonumber\\
&& \hskip 103pt + 2 {\hat k}_z {\hat p}_z (\hat{\bm k}\cdot\hat{\bm p}) \sum_{\beta} F_k^{+\beta} F_{k-q}^{-\beta} F_{k-Q}^{+\beta}
     \sum_{\beta'} F_p^{+\beta'} F_{p-q}^{-\beta'} F_{p-Q}^{+\beta'} 
\nonumber\\
&& \hskip 80pt + 2{\hat k}_z^2 \left[1+(\bm{\hat k}\cdot\bm{\hat p})^2\right] 
       \sum_{\beta} F_k^{+\beta} F_k^{+\beta} F_{k-q}^{-\beta} F_{k-Q}^{+\beta}        
       \sum_{\beta'} F_p^{+\beta'} F_{p-q}^{-\beta'}
\nonumber\\
&& \hskip 110pt +4 {\hat k}_z^2 (\hat{\bm k}\cdot\hat{\bm p}) \sum_{\beta} \beta 
                    F_k^{+\beta} F_k^{+\beta} F_{k-q}^{-\beta} F_{k-Q}^{+\beta} 
                           \sum_{\beta'} \beta' F_p^{+\beta'} F_{p-q}^{-\beta'} \biggr\}
\nonumber\\
&& \hskip 125pt +\ (\text{the same terms with ${\bm Q}\to -{\bm Q}$})
\label{eq:4.2b}
\eea
\ese
\end{widetext}

Power counting shows that all of the integrals in Eq.~(\ref{eq:4.2b}) scale as a constant plus $\vert{\bm Q}\vert^{d-1}$ for spatial dimensions
$d<3$, and as ${\bm Q}^2 \ln\vert{\bm Q}\vert$ for $d=3$, which is the expected nonanalytic behavior. Determining the prefactor
for $d<3$ would be hard. However, for $d=3$, and to leading logarithmic accuracy, we can expand the integrands
in powers of ${\bm Q}$ to $O({\bm Q}^2)$, and replace the resulting logarithmic divergence with $\ln\vert{\bm Q}\vert$. 
After lengthy calculations we obtain, for $h=T=0$ and $v\ll\vF$,
\bea
\delta\chisL({\bm Q}) &=& \frac{1}{30}\,(\NF\Gamma_{\text{t},3})^2 (\vF{\bm Q})^2 \frac{1}{V}\sum_{\bm q} \frac{1}{(\vF\vert{\bm q}\vert)^3}
\nonumber\\
&=& \NF\,(\NF\Gamma_{\text{t},3})^2\frac{1}{30}\, \left({\bm Q}/\kF\right)^2 \ln(\kF/\vert{\bm Q}\vert)\qquad
\label{eq:4.3}
\eea
where we have omitted the constant contribution.

While this result has the same functional form as the corresponding one in a LFL,\cite{Belitz_Kirkpatrick_Vojta_1997}
we stress that the underlying soft-mode structure is very different: The LFL soft modes that were responsible for the
nonanalytic behavior in the absence of a spin-orbit interaction have become massive, see Eq.~(\ref{eq:3.6a}), and the
soft modes that are responsible for the present result mix chiralities, which is why only the interaction amplitude
$\Gamma_{\text{t},3}$ contributes. We also note that our result is independent of $v$ and nonzero in the limit $v\ll\vF$.
This does {\em not} imply that it remains unchanged in the true $v\to 0$ limit: We have neglected terms that have
become massive due to $v\neq 0$, so we can no longer take the $v\to 0$ limit. 
If one keeps the terms that are soft for $v=0$, but massive for $v>0$, then one finds that in the limit
$v\to 0$ they cancel the contribution shown in Eq.~(\ref{eq:4.3}), leaving behind the known result for a LFL.\cite{Belitz_Kirkpatrick_Vojta_1997}
We will come back to these points in Sec.~\ref{sec:V}. 

\subsubsection{Temperature dependence}
\label{subsubsec:IV.A.2}

We now turn to the temperature dependence of $\delta\chisL$ at ${\bm Q} = h = 0$. For ${\bm Q}=0$ it is possible to
identically rewrite all of the terms in Eq.~(\ref{eq:4.2b}) as products of convolutions of three Green functions. This
can be achieved by means of partial integrations with respect to the $z$-component of the hydrodynamic wave vector
${\bm q}$. We find
\begin{widetext}
\bse
\label{eqs:4.4}
\bea
\delta\chisL(T) &=& -2 (\Gamma_{t,3})^2 \frac{\vF^2}{\vF^2 - v^2} \sum_{q}{}^{'}\sum_{k,p} 
    {\hat k}_z {\hat p}_z (1 - \hat{\bm k}\cdot\hat{\bm p})^2 \left(F_k^{++}\right)^2 F_{k-q}^{-+}\,\left(F_p^{+-}\right)^2 F_{p-q}^{--}
\nonumber\\
&=& -2 (\Gamma_{t,3})^2 \frac{\vF^2}{\vF^2 - v^2} \sum_{q}{}^{'} \left[  
        -2\sum_{i=1}^3 {\cal F}_{zi}^{(2)+}(q) {\cal F}_{zi}^{(2)-}(q) +\sum_{i,j=1}^3 {\cal F}_{zij}^{(3)+}(q) {\cal F}_{zij}^{(3)-}(q)  \right]
\label{eq:4.4a}
\eea
where
\be
{\cal F}_z^{(1)\beta}(q) = \sum_k {\hat k}_z \left(F_k^{+\beta}\right)^2 F_{k-q}^{-\beta} \quad,\quad
{\cal F}_{zi}^{(2)\beta}(q) = \sum_k {\hat k}_z {\hat k}_i \left(F_k^{+\beta}\right)^2 F_{k-q}^{-\beta} \quad,\quad
{\cal F}_{zij}^{(3)\beta}(q) = \sum_k {\hat k}_z {\hat k}_i {\hat k}_{j}\left(F_k^{+\beta}\right)^2 F_{k-q}^{-\beta}
\label{eq:4.4b}
\ee
\ese
We now use a spectral representation for the $\cal F$ to perform the summation over the hydrodynamic frequency.
This yields
\bea
\delta\chisL(T) &=& 2 (\Gamma_{t,3})^2 \frac{\vF^2}{\vF^2 - v^2}\,\frac{1}{V}\sum_{\bm q} \int_{-\infty}^{\infty} \frac{dx}{\pi}\,\frac{dy}{\pi}
     \, \frac{n(x/T) - n(y/T)}{x - y}\, \left[ {\cal F}_z^{(1)+}{}^"(q) {\cal F}_z^{(1)-}{}^"(q) \right.
\nonumber\\
&& \hskip 150pt  - 2\sum_{i=1}^3 {\cal F}_{zi}^{(2)+}{}^"(q) {\cal F}_{zi}^{(2)-}{}^"(q) 
                           + \left. \sum_{i,j=1}^3 {\cal F}_{zij}^{(3)+}{}^"(q) {\cal F}_{zij}^{(3)-}{}^"(q) \right]
\label{eq:4.5}                        
\eea
\end{widetext}
Here the ${\cal F}{}^{\,"}$ are the spectra of the $\cal F$, which contain a continuous part as well as delta-function
contributions, and $n(x) = 1/(e^x-1)$ is the Bose distribution function. The low-$T$ behavior can now be determined
by standard asymptotic analysis. Power counting shows that the leading singularity in $d<3$ spatial dimensions is
$T^{d-1}$, and $T^2\ln T$ in $d=3$, as expected. Focusing on $d=3$ again, structural considerations show that
${\cal F}^{(2)}$ does not contribute to the singular behavior, while ${\cal F}^{(1)}$ and ${\cal F}^{(3)}$ do. The
contributions from ${\cal F}^{(1)}$ cancel internally, and those from ${\cal F}^{(3)}$ yield the final result
\be
\delta\chisL(T) = \NF\,(\NF\Gamma_{t,3})^2\,\frac{\pi^4}{48}\,\left(T/\epsilonF\right)^2 \ln(\epsilonF/T)
\label{eq:4.6}
\ee

As expected from the general arguments in Sec.~\ref{sec:III}, there is a nonanalytic temperature dependence
whose functional form and sign are the same as for the ${\bm Q}$-dependence at $T=0$. 
While the result is formally independent of $v$, the comments made after Eq.~(\ref{eq:4.3}) apply
here as well: We have neglected modes that are massive for $v>0$. If kept, they will cancel the term shown in
the limit $v\to 0$. In contrast to the case of the $Q$-dependence, here the cancellation is exact, and the prefactor
of the $T^2\ln T$ term in a LFL is zero, at least to second order in the interaction.\cite{Carneiro_Pethick_1977, Belitz_Kirkpatrick_Vojta_1997}
The nonzero result in a DFL
%
underscores the
point that the absence of
a $T^2\ln T$ term in a  in a LFL is not of structural significance, but due to a prefactor that is accidentally zero
in $d=3$, see the discussion in Sec.~\ref{sec:V}.

\subsubsection{Magnetic-field dependence}
\label{subsubsec:IV.A.3}

Finally, we consider the dependence on the magnetic field $h$ at ${\bm Q} = T = 0$. Since ${\bm Q}=0$, we can
write $\delta\chisL$ in the form of Eqs.~(\ref{eqs:4.4}). In order to extract the leading nonanalyticity in $d=3$ we
expand the integrand to $O(h^2)$ and realize that the resulting logarithmic divergence is cut off by $h$. The 
leading singular behavior is obtained from terms that maximize the number of Green functions in the
integrands. This justifies the approximation we made in Sec.~\ref{subsubsec:II.B.2}, where we neglected all
occurrences of $h$ in the numerator of the Green function. To leading logarithmic accuracy we obtain
\bea
\delta\chisL(h) &=& \frac{8}{5}\,(\NF\Gamma_{t,3})^2\,h^2 \frac{1}{V}\sum_{\bm q} \frac{1}{(\vF\vert{\bm q}\vert)^3}
\nonumber\\
                        &=& \NF (\NF\Gamma_{t,3})^2\, \frac{2}{5}\, (h/\epsilonF)^2 \ln(\epsilonF/h)
\label{eq:4.7}
\eea
Again, this is consistent with the prediction based on the general arguments in Sec.~\ref{sec:III}. As explained
there, this is, in a well defined sense, the most fundamental nonanalyticity, since $h$ is the field conjugate to
the observable whose susceptibility we are calculating, viz., the magnetization. The nonanalytic dependences 
on ${\bm Q}$ and $T$ follow from the one on $h$ due to scaling.

\subsection{Density susceptibility, and related correlation functions}
\label{subsec:IV.B}

We next ascertain that the density susceptibility $\chi_{\text{n}}$ has no nonanalytic behavior, as predicted by the general arguments
of Sec.~\ref{sec:III}. To second order in the interaction the diagrams are the same as in Fig.~\ref{fig:4}, but the external
vertices now carry a unit matrix in spin space rather than a $\sigma_3$, which changes the result of performing the spin
traces. The realizations of the diagrams in terms of $\Gamma_{\mu,1}$ and $\Gamma_{\mu,2}$ vanish for the same
symmetry reasons as in the case of $\chisL$. For the realizations in terms of $\Gamma_{\mu,3}$ we find
\begin{widetext}
\bse
\label{eqs:4.8}
\bea
\delta\chi_{\text{n}}({\bm Q}) &=& \frac{1}{4}\,(\Gamma_{t,3})^2 \sum_{q}{}^{'}\sum_{k,p} 
     \biggl\{ \left[1+(\bm{\hat k}\cdot\bm{\hat p})^2\right] \sum_{\beta} F_k^{+\beta} F_{k-q}^{-\beta} F_{k-Q}^{+\beta}
     \sum_{\beta'} \left(F_p^{+\beta'} F_{p-q}^{-\beta'} F_{p-Q}^{+\beta'} + F_p^{-\beta'} F_{p+q}^{+\beta'} F_{p+Q}^{-\beta'}\right)
\nonumber\\
&& \hskip 103pt + 2 (\hat{\bm k}\cdot\hat{\bm p}) \sum_{\beta} \beta F_k^{+\beta} F_{k-q}^{-\beta} F_{k-Q}^{+\beta}
     \sum_{\beta'} \beta' \left(F_p^{+\beta'} F_{p-q}^{-\beta'} F_{p-Q}^{+\beta'} + F_p^{-\beta'} F_{p+q}^{+\beta'} F_{p+Q}^{-\beta'}\right)
\nonumber\\
&& \hskip 40pt +  \left[1+(\bm{\hat k}\cdot\bm{\hat p})^2\right] 
       \sum_{\beta} \left( 2 F_k^{+\beta} F_k^{+\beta} F_{k-q}^{-\beta} F_{k-Q}^{+\beta} + F_k^{+\beta} F_{k-q}^{-\beta} F_{k-Q}^{+\beta} F_{k-q-Q}^{-\beta} \right)          
       \sum_{\beta'} F_p^{+\beta'} F_{p-q}^{-\beta'}
\nonumber\\
&& \hskip 65pt +2  (\hat{\bm k}\cdot\hat{\bm p}) \sum_{\beta} \beta 
                    \left( 2 F_k^{+\beta} F_k^{+\beta} F_{k-q}^{-\beta} F_{k-Q}^{+\beta} +F_k^{+\beta} F_{k-q}^{-\beta} F_{k-Q}^{+\beta} F_{k-q-Q}^{-\beta}\right)
                           \sum_{\beta'} \beta' F_p^{+\beta'} F_{p-q}^{-\beta'} \biggr\}
\nonumber\\
&& \hskip 125pt +\ (\text{the same terms with ${\bm Q}\to -{\bm Q}$})
\label{eq:4.8a}\\
\nonumber\\
&=& 0
\label{eq:4.8b}
\eea
\ese
\end{widetext}
As in the case of $\delta\chisL$, 
the expression in terms of integrals, Eq.~(\ref{eq:4.8a}),
is valid only for the leading nonanalytic contributions to $\delta\chi_{\text{n}}$.
We see that Eq.~(\ref{eq:4.8a}) is very similar to the corresponding result for $\delta\chisL$,
Eq.~(\ref{eq:4.2a}). However, the symmetry relations from Eqs.~(\ref{eq:C.2b}) and (\ref{eq:C.4c}) now lead to terms being equal and opposite 
that were equal in the case of $\chisL$. As a result, the contributions proportional to $\Gamma_{\mu,3}^2$ cancel as well, as they 
must according to the general arguments. We conclude that the density susceptibility $\chi_{\text{n}}$ is an analytic function of the
wave number, the temperature, and the magnetic field. Consistent with the very general reasons underlying this result, the
cancellation occurs at a structural level, and it is not necessary to evaluate any of the integrals in order to see it.

The symmetry relations that lead to this null result are independent of the angular factors encoded in the function $f(\hat{\bm k})$
in Eqs.~(\ref{eqs:C.1}) - (\ref{eqs:C.4}). As a result, the absence of nonanalyticities also applies to the density current
susceptibility and to all high-rank tensor susceptibilities related to the density. This is important for the stability of Fermi
liquids, as we explain in Sec.~\ref{sec:V}. 

\subsection{Density of states}
\label{subsec:IV.C}

The density of states (DOS) $N(\omega)$, with $\omega$ the difference in energy space from the Fermi surface, 
is given by the trace of the Green function via
\bse
\label{eqs:4.9}
\be
N(\omega) = \text{Re}\,Q(i\omega \to \omega\to i0)
\label{eq:4.9a}
\ee
where
\be
Q(i\omega_n) = \frac{i}{\pi}\,\frac{1}{V} \sum_{\bm k} \tr\ G_n({\bm k})
\label{eq:4.9b}
\ee
\ese
\begin{figure}[b]
\includegraphics[width=3.7cm]{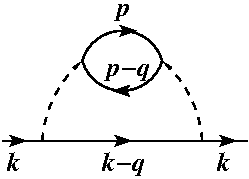}
\caption{Contribution to the density of states at second order in the interaction amplitudes. The notation is the same as in Fig.~\ref{fig:4}.}
\label{fig:5}
\end{figure}
For noninteracting electrons, $N(\omega\to 0) = 4\NF$. We are interested in corrections $\delta N(\omega)$ that are nonanalytic functions
of the frequency $\omega$ due to the effects of the soft modes. To second order in the interaction there is only one diagram that is of
one-loop type in the sense of Ref.~\onlinecite{two-loop_footnote}, which is shown in Fig.~\ref{fig:5}. 
Realizing this diagram in all
possible ways with the interaction amplitudes defined in Eq.~(\ref{eq:2.15}), and keeping only terms with hydrodynamic content, we find seven 
different contributions. In terms of convolutions of the Green function $F$ defined in Eq.~(\ref{eq:2.10c}) these are

\begin{widetext}
\bse
\label{eqs:4.10}
\bea
\delta Q_{\text{ss},1}(i\omega) &=& \frac{-i}{\pi}\,\left(\Gamma_{\text{s},1}\right)^2 \sum_{\alpha} \sum_{q}{}^{'} \frac{1}{V}\sum_{\bm k} \sum_{\beta}
   F_k^{\alpha\beta} F_k^{\alpha\beta} F_{k-q}^{\alpha\beta} \sum_p \sum_{\beta'} F_p^{\alpha\beta'} F_{p-q}^{\alpha\beta'}
\label{eq:4.10a}\\
\delta Q_{\text{st},1}(i\omega) &=& \frac{-2i}{\pi}\,\Gamma_{\text{s},1} \Gamma_{\text{t},1} \sum_{\alpha} \sum_{q}{}^{'} \frac{1}{V}\sum_{\bm k}
   (\hat{\bm q}\cdot\hat{\bm k}) \sum_{\beta} \beta
   F_k^{\alpha\beta} F_k^{\alpha\beta} F_{k-q}^{\alpha\beta} \sum_p (\hat{\bm q}\cdot\hat{\bm p})\sum_{\beta'} \beta' F_p^{\alpha\beta'} F_{p-q}^{\alpha\beta'}
\label{eq:4.10b}\\
\delta Q_{\text{tt},1}(i\omega) &=& \frac{-i}{\pi}\,\left(\Gamma_{\text{t},1}\right)^2 \sum_{\alpha} \sum_{q}{}^{'} \frac{1}{V}\sum_{\bm k} 
   (\hat{\bm q}\cdot\hat{\bm k})^2 \sum_{\beta}
   F_k^{\alpha\beta} F_k^{\alpha\beta} F_{k-q}^{\alpha\beta} \sum_p (\hat{\bm q}\cdot\hat{\bm p})^2\sum_{\beta'} F_p^{\alpha\beta'} F_{p-q}^{\alpha\beta'}
\label{eq:4.10c}\\
\delta Q_{\text{ss},2}(i\omega) &=& \frac{-i}{\pi}\,\left(\Gamma_{\text{s},2}\right)^2 \sum_{\alpha\neq\alpha'} \sum_{q}{}^{'} \frac{1}{V}\sum_{\bm k} \sum_{\beta}
   F_k^{\alpha\beta} F_k^{\alpha\beta} F_{k-q}^{\alpha\beta} \sum_p \sum_{\beta'} F_p^{\alpha'\beta'} F_{p-q}^{\alpha'\beta'}
\label{eq:4.10d}\\
\delta Q_{\text{st},2}(i\omega) &=& \frac{-2i}{\pi}\,\Gamma_{\text{s},2} \Gamma_{\text{t},2} \sum_{\alpha\neq\alpha'} \sum_{q}{}^{'} \frac{1}{V}\sum_{\bm k}
   (\hat{\bm q}\cdot\hat{\bm k}) \sum_{\beta} \beta
   F_k^{\alpha\beta} F_k^{\alpha\beta} F_{k-q}^{\alpha\beta} \sum_p (\hat{\bm q}\cdot\hat{\bm p})\sum_{\beta'} \beta' F_p^{\alpha'\beta'} F_{p-q}^{\alpha'\beta'}
\label{eq:4.10e}\\
\delta Q_{\text{tt},2}(i\omega) &=& \frac{-i}{\pi}\,\left(\Gamma_{\text{t},2}\right)^2 \sum_{\alpha\neq\alpha'} \sum_{q}{}^{'} \frac{1}{V}\sum_{\bm k} 
   (\hat{\bm q}\cdot\hat{\bm k})^2 \sum_{\beta}
   F_k^{\alpha\beta} F_k^{\alpha\beta} F_{k-q}^{\alpha\beta} \sum_p (\hat{\bm q}\cdot\hat{\bm p})^2\sum_{\beta'} F_p^{\alpha'\beta'} F_{p-q}^{\alpha'\beta'}
\label{eq:4.10f}\\
\delta Q_{\text{tt},3}(i\omega) &=& \frac{-i}{\pi}\,\left(\Gamma_{\text{t},3}\right)^2 \sum_{\alpha\neq\alpha'} \sum_{q}{}^{'} \left[
   \frac{1}{V}\sum_{\bm k} \sum_{\beta} F_k^{\alpha\beta} F_k^{\alpha\beta} F_{k-q}^{\alpha'\beta} \sum_p 
        \sum_{\beta'} F_p^{\alpha\beta'} F_{p-q}^{\alpha'\beta'} \right.
   \nonumber\\
   &&\hskip 95pt + \frac{2}{V} \sum_{\bm k} (\hat{\bm q}\cdot\hat{\bm k}) \sum_{\beta} F_k^{\alpha\beta} F_k^{\alpha\beta} F_{k-q}^{\alpha'\beta} \sum_p 
        \sum_{\beta'} F_p^{\alpha\beta'} F_{p-q}^{\alpha'\beta'}
    \nonumber\\
    &&\hskip 95pt  \left. +\frac{1}{V}\sum_{\bm k} (\hat{\bm q}\cdot\hat{\bm k})^2 \sum_{\beta} F_k^{\alpha\beta} F_k^{\alpha\beta} F_{k-q}^{\alpha'\beta} \sum_p 
        (\hat{\bm q}\cdot\hat{\bm p})^2 \sum_{\beta'} F_p^{\alpha\beta'} F_{p-q}^{\alpha'\beta'} \right]
\label{eq:4.10g}
\eea
\ese

Power counting shows that these integrals scale as a constant plus $\vert\omega\vert^{d-1}$ for spatial dimensions $1<d<3$, and as
$\omega^2\ln\vert\omega\vert$ in $d=3$, which is the behavior expected from the general arguments in Sec.~\ref{subsubsec:III.D.1}.
An explicit calculation shows that the nonanalyticity is absent for $d=3$ (but {\em not} for $d<3$) in $\delta Q_{\text{tt},1}$,
$\delta Q_{\text{tt},2}$, and the third contribution to $\delta Q_{\text{tt},3}$.\cite{higher_angular_momenta_footnote} We obtain
\bse
\label{eqs:4.11}
\bea
\delta N(\omega) &=& 4\NF\left\{\left[\left(\gamma_{\text{s},1}\right)^2 + \left(\gamma_{\text{s},2}\right)^2 + \left(\gamma_{\text{s},3}\right)^2 \right] f_0(v/\vF)\right.
\nonumber\\
&& \hskip 30pt + \left.\left[\gamma_{\text{s},1}\gamma_{\text{t},1} + \gamma_{\text{s},2}\gamma_{\text{t},2} + \left(\gamma_{\text{t},3}\right)^2\right]
     f_1(v/\vF) \right\} \frac{1}{32}\left(\frac{\omega}{\epsilonF}\right)^2 \ln(\vert\omega\vert/\epsilonF) + O(\omega^2)
\label{eq:4.11a}
\eea
where
$\gamma_a = 4\NF\Gamma_a$ ($a = \text{s},1$ etc.), and
\be
f_0(x) = (1+x^2)/(1-x^2)^3\quad,\quad f_1(x) = -2 x^2/(1 - x^2)^3
\label{eq:4.11b}
\ee
\ese
\end{widetext}

On the Fermi surface, $\omega=0$, we find an analogous result for $\delta N(T)$ as a function of the temperature: The result
has the same form as Eq.~(\ref{eq:4.11a}) with $\omega^2 \ln\vert\omega\vert$ replaced by $T^2\ln T$, only the numerical
prefactor is slightly different. 

A nonzero magnetic field affects only $\delta Q_{\text{tt},3}$, which is the only contribution containing convolutions that mix
chiralities. Accordingly, the term proportional to $(\gamma_{\text{t},3})^2 \omega^2\ln\vert\omega\vert$ gets replaced,
for $h > \omega,T$, by $(\gamma_{\text{t},3})^2 h^2\ln h$, again with a different numerical prefactor. The other contributions remain unaffected.

\section{DIscussion}
\label{sec:V}

We conclude by discussing various aspects of our results, and by elaborating on some points we have
mentioned in passing only.

\subsection{Different types of Fermi liquids}
\label{subsec:V.A}

\subsubsection{Landau Fermi liquids versus Dirac Fermi liquids}
\label{subsubsec:V.A.1}

A Fermi liquid in general is characterized by the existence of well-defined quasiparticles that are continuously related
to the excitations in a Fermi gas. As we have mentioned in the Introduction, the action given by Eq.~(\ref{eq:2.16}), with
or without the spin-orbit coupling $v$, does indeed describe a Fermi liquid. This can be seen explicitly by considering
the electron self energy, which is given by the diagram in Fig.~\ref{fig:5} without the two external Green functions. If
we average over the Fermi surface, we obtain the expressions in Eqs.~(\ref{eqs:4.10}) with the factor $F_k^{\alpha\beta} F_k^{\alpha\beta}$
common to all of them replaced by a delta function $\delta(\xi_{\bm k})$ that pins ${\bm k}$ to the Fermi surface. Given that the DOS
scales as $\omega^{d-1}$, it is obvious that the self energy scales as $\omega^d$, i.e., well-defined quasiparticles
exist for all dimensions $d>1$. 

However, the nature of the Fermi liquid is drastically different depending on whether or not the spin-orbit coupling $v$
is present. As illustrated in Eqs.~(\ref{eqs:3.5}, \ref{eqs:3.6}), the modes that are soft of the first kind with respect to the
magnetic field $h$ for $v=0$ are no longer soft for $v>0$. This is related to the fact that the spin-orbit coupling destroys
spin conservation. The modes that are soft instead are excitations that mix chiralities, see Eq.~(\ref{eq:3.6b}), a concept
that disappears in the limit $v\to 0$ when the two chirality channels become degenerate. For our purposes, this is the
crucial difference between a Landau Fermi liquid (LFL) at $v=0$ and a Dirac Fermi liquid (DFL) at $v>0$. As noted
in connection with Eq.~(\ref{eq:3.7}), the soft excitations in a DFL have a structure that is qualitatively different from
that of density fluctuations. In this sense they are as different from the soft modes in a LFL as particle-particle or
Cooper-channel excitations are from particle-hole excitations. 

\subsubsection{Effects of the gap $\Delta$, and the limit of large $v$}
\label{subsubsec:V.A.2}

We still need to show that our results are not qualitatively altered if one considers the general model with $\Delta\neq 0$.
As can be seen from Eq.~(\ref{eq:2.9}), the gap affects the Green function in two distinct ways. A trivial effect is that
the expression $(\alpha v{\bm k} - {\bm h})^2$ in the denominator of the Green function gets modified according to
\bea
(\alpha v{\bm k} - {\bm h})^2 &=& v^2{\bm k}^2 - \alpha 2 v k_z h + h^2
\nonumber\\
&\rightarrow& v^2{\bm k}^2 + \Delta^2 -\alpha 2 h \sqrt{v^2 k_z^2 + \Delta^2} + h^2
\nonumber\\
\label{eq:5.1}
\eea
which modifies $F_k^{\alpha\beta}$, Eq.~(\ref{eq:2.10c}). A less trivial effect is that the Green function is no
longer diagonal in the chirality index due the term $(\pi_1\otimes\sigma_0)\Delta$ in the numerator of
Eq.~(\ref{eq:2.9}). This allows for realizations of the diagrams in Fig.~\ref{fig:4} that are not possible if
$\Delta = 0$. However, in order for convolutions of the $F$ to be soft of the first kind with respect to the
magnetic field they still need to mix chiralities. This can now be achieved by means of the Green functions
rather than the interaction amplitudes, but the soft-mode structure remains unchanged and  the integrals are 
still the same as in Eqs.~(\ref{eqs:4.2}) (except for the modified $F$). The only effect of the gap is therefore to 
produce small corrections of $O((\Delta/v\kF)^2)$ to the results we obtained in Sec.~\ref{sec:IV}.

Our explicit results have been obtained in the limit $v\ll v_0$, with $v_0$ the atomic-scale velocity.
An obvious question is what happens in the limit $v/v_0\alt 1$. Consider the spin susceptibility
$\chis$ as an example. The general considerations in
Sec.~\ref{sec:III} show that the soft modes of the first kind with respect to the magnetic field are given
by convolutions with all branch indices $\beta$ the same, while the chirality indices must not be all the
same. Furthermore, in order for the magnetic field dependence to not disappear upon a simple shift
of the wave number, the two soft modes in any one-loop contribution to $\chis$ must carry different
branch indices. This is consistent with the structure of the branch and chirality indices in, e.g.,
Eq.~(\ref{eq:4.4a}). As a result, the interaction couples a branch $E^{1\beta}$ of the single-particle
specrum, Eqs.~(\ref{eqs:2.3}), with a branch $E^{2\beta}$. However, one of these two branches will
not contribute to the Fermi surface in the limit of large $v$, as can be seen from Fig.~\ref{fig:2}.
This suggests that in this limit the spin susceptibility does not have a nonanalyticity in systems
with $\Delta=0$.  $\Delta\neq 0$, on the other hand, will eliminate the condition that the two
soft modes must carry different branch indices and will likely restore the nonanalyticity, albeit
with a prefactor that is proportional to $\Delta$. These remarks apply to the spin susceptibility,
other observables will behave differently. For instance, for the density of states all of the soft
modes are of the first kind. The restrictions that are in place for $\chis$ therefore do not apply,
and we expect the behavior for large $v$ to be qualitatively the same as for small $v$.
These qualitative considerations suggest that the
behavior in the limit of large $v$ is quite complex, and different in various aspects from that for
small $v$. This limit requires a separate detailed investigation.

\subsubsection{Different types of Dirac Fermi liquids}
\label{subsubsec:V.A.3}

A linear spectrum at small momenta is not sufficient to uniquely define a Dirac Fermi liquid. This can be
seen by comparing the model for a Dirac metal we have discussed, where the Pauli matrices in the
$v\bm\sigma\cdot{\bm k}$ term represents the physical spin degree of freedom, with the graphene-type models mentioned
after Eq.~(\ref{eq:2.1a}), where they represent a pseudo-spin. In the former case, the discussion in
Sec.~\ref{subsubsec:V.A.1} applies. In a graphene-type model, on the
other hand, the situation is different. Since the Pauli matrices in the $v\bm\sigma\cdot{\bm k}$ term and in the Zeeman term
represent different degrees of freedom, the soft modes of the first kind with respect to the magnetic field are unaffected
by the coupling constant $v$, and the physics behind the nonanalytic spin susceptibility the same as in a LFL. We note,
however, that these remarks do not pertain to, e.g., the density of states. Since the nonanalyticity of the latter relies on
modes that are soft with respect to the frequency-dependent chemical potential introduced in Eq.~(\ref{eq:3.9}), rather
than the magnetic field, the physical difference between the two models does not apply.

We also note that other types of spin-orbit interactions exist, which leads to different types of single-particle spectra.
For instance, the touching point may be quadratic rather than linear.\cite{Abrikosov_Beneslavskii_1970} For some of
these systems the behavior in the metallic regime will likely be the same as for the Dirac metals discussed here. 
For a two-dimensional electron system with a Rashba spin-orbit coupling the analog of Landau
Fermi-liquid theory has been developed in Ref.~\onlinecite{Ashrafi_Rashba_Maslov_2013}, which also calculated the
effective mass and various thermodynamic quantities in terms of Fermi-liquid parameters. A similar study for the
two-dimensional helical Fermi liquid on the surface of certain bulk topological insulators was carried out in
Ref.~\onlinecite{Lundgren_Maciejko_2015}.
The spin susceptibility for a two-dimensional electron system with a spin-orbit interaction of Rashba type was
considered in Ref.~\onlinecite{Zak_Maslov_Loss_2010}, and the consequences for nuclear spin polarization were
discussed in Ref.~\onlinecite{Zak_Maslov_Loss_2012}. Our result that the chirality-independent modes, 
Eq.~(\ref{eq:3.6a}), become massive as a result of the spin-orbit interaction, are consistent with the results of
these papers. However, we find that as a result of the chirality degrees of freedom there are different soft modes
that remain soft for $v>0$, see Eq.~(\ref{eq:3.6b}). It is the existence of the latter that leads to a DFL having
nonanalyticities very similar to those in a LFL, although as a result of a different set of soft modes.

\subsection{Nature of the soft modes}
\label{subsec:V.B}

As we have discussed in Sec.~\ref{sec:III}, the soft modes that underly all of the nonanalyticities we have derived
are phase-space four-fermion correlation functions that are local in frequency space and of the general form
\bea
&&\langle {\bar\psi}_{n_1+m}({\bm k}+{\bm q}/2)\,\psi_{n_1}({\bm k}-{\bm q}/2)
\nonumber\\
&&\qquad\qquad\times {\bar\psi}_{n_2-m}({\bm p}-{\bm q}/2)\,\psi_{n_2}({\bm p}+{\bm q}/2)\rangle\qquad
\label{eq:5.2}
\eea
For noninteracting electrons, the correlation factorizes into a product of two Green functions and we obtain, in the LFL case,
Eq.~(\ref{eq:3.1}), but at $T=0$ it remains soft even in the case of interactions.\cite{Belitz_Kirkpatrick_2012a} Their
physical interpretation is in terms of the Goldstone modes of a spontaneously broken rotational symmetry in frequency
space, see Appendix~\ref{app:B}. In calculations that are perturbative with respect to the interaction they enter via convolutions 
of noninteracting Green functions, the most basic one for a DFL is given in Eqs.~(\ref{eqs:3.4}). We note that these Goldstone
modes do not describe density fluctuations: The latter are local in time space, whereas the Goldstone modes are local in
frequency space. This distinction gets blurred in the case of noninteracting electrons (e.g.,
Eq.~(\ref{eq:3.5b}) is just the hydrodynamic part of the Lindhard function, which does describe density fluctuations),
but it is physically important in general. For instance, the physical density susceptibility in an interacting system has
no dephasing rate, and remains soft, as a result of particle number conservation. This is not true for the Goldstone
modes, which do acquire a mass at nonzero temperature. In a DFL, the chirality structure invalidates 
the interpretation as density fluctuations of even the noninteracting reference system, see the discussion after Eq.~(\ref{eq:3.6b})

As the discussion in Sec.~\ref{subsec:III.D} has shown, the properties of the soft modes in the presence of the
external field conjugate to the observable under consideration are crucial. Specifically, only modes that are soft of
the first kind with respect to the conjugate field, i.e., acquire a small mass upon the application of a small field,
can lead to nonanalyticities of the corresponding variable. This observation also underscores the fact that the
fundamental nonlinear behavior of any susceptibility is the one with respect to the relevant conjugate field. The
nonanalytic dependences on anything else (e.g., wave number, frequency, or any fields other than the conjugate
one) are derivative and follow from how these parameters scale with the conjugate field. As an example, the
nonanalytic dependence of the spin susceptibility on the magnetic field $h$ expressed in Eq.~(\ref{eq:4.7}) is the
fundamental one for this observable. The nonanalyticities as a function of the temperature, Eq.~(\ref{eq:4.6}), and
the wave number, Eq.~(\ref{eq:4.3}), are consequences of the fact that both the temperature and the wave
number scale linearly with $h$. 

Conversely, without  a mode that is soft of the first kind with respect to the respective conjugate field there can
be no nonanalytic dependence on any parameter. An example is the absence of a nonanalyticity in the density
susceptibility and related correlation functions that is predicted by the arguments in Sec.~\ref{subsec:III.D} and 
confirmed by the calculation in Sec.~\ref{subsec:IV.B}. This null result is of great physical importance: Since
the diamagnetic susceptibility is related to the coefficient of the ${\bm Q}^2$ term in the wave-number dependent
number-density current suceptibility,\cite{Pines_Nozieres_1989} a nonanalyticity in the density channel analogous
to the one in the spin-density channel would imply ideal diamagnetism.

Another consequence of the structure of the soft modes, and their dependence on various parameters, is the
fact that the results for the nonanalyticity in the spin susceptibility, Eqs.~(\ref{eq:4.3}, \ref{eq:4.6}, \ref{eq:4.7}) 
are nonzero in the limit $v\to 0$. This does not contradict the fact that the spin-orbit coupling is necessary in
order to obtain these terms in the first place. The point is that other modes have acquired a mass due to the
spin-orbit interaction and therefore do not contribute to any nonanalytic behavior. If one kept all contributions
to the susceptibility, from massive modes as well as from massless ones, then some of the terms that are made
massive by the spin-orbit interaction would, in the limit $v\to 0$, cancel the terms we have calculated, leaving behind the result for
a LFL. As written, our results for the spin susceptibility are valid for $\Delta,h \ll v\kF \ll \epsilonF$. Corrections of 
order $\Delta/v\kF$ and $v/\vF$ can be calculated if desired, they just make the prefactors of the nonanalyticities 
more complicated. In our calculation of the density of states, Sec.~\ref{subsec:IV.C}, we have relaxed the
condition $v\kF \ll \epsilonF$ since in this case some contributions vanish as $v\to 0$.

\subsection{Importance of the soft-mode structure}
\label{subsec:V.C}

An important conceptual point is that the general considerations in Sec.~\ref{sec:III} suffice for determining the
universal nonanalytic behavior of any observable. No explicit calculations are necessary unless one wants to
determine the prefactor of the nonanalyticity, which is model dependent and hence non-universal in any case. 
The calculations in Sec.~\ref{sec:IV} merely demonstrate that the results of explicit calculations are consistent
with the general arguments, as they must be. It is also worth noting that the explicit calculations for the DFL
are very involved, and guidance from the general structural arguments is very helpful for the asymptotic analysis
needed to extract the leading nonanalytic behavior.

The structural arguments can be taken to a higher level by employing field-theoretic and renormalization-group (RG)
techniques. In fact, some elements of such an analysis we already employed in Sec.~\ref{sec:IV} by restricting
the calculation to diagrams that correspond to one-loop terms in a field-theoretic description. More generally,
a LFL is associated with a stable renormalization-group (RG) fixed point that characterizes the entire LFL phase
(as opposed to a critical RG fixed point that characterizes a phase transition, see Ref.~\onlinecite{Ma_1976} for a
discussion of this important distinction). Such a fixed point has been identified in both a fermionic field theory\cite{Shankar_1994}
and an effective bosonic one.\cite{Belitz_Kirkpatrick_2012a} Our soft-mode analysis in Sec.~\ref{sec:III} implies that the
spin-orbit interaction is a relevant operator with respect to this fixed point and hence makes it unstable. In addition, there must
be another unstable fixed point, one describing a Dirac semimetal, with respect to which the chemical potential is a relevant
operator. In the presence of both the spin-orbit interaction and a nonzero chemical potential the RG flow must lead from
either unstable fixed point to a stable one that describes a DFL.  In such a description, the leading nonanalyticities we have derived 
will be associated with the least irrelevant operators at the stable fixed point. Such a description will do much more than just
confirm the perturbative results. Since the nature the nonanalyticities is tied to the scale dimensions of the least irrelevant
operators, it can provide a proof of the hypothesis that the perturbative results are exact as far as the exponents of the nonanalyticities are
concerned. Higher orders in perturbation theory and/or a loop expansion will change the prefactor, but cannot change
the exponents. This is impossible to ascertain within perturbation theory. It has been shown to be true for a LFL,\cite{Belitz_Kirkpatrick_2014}
and it is important to establish the corresponding result for a DFL. Such an analysis will also underscore the point that
the DFL is qualitatively different from a LFL, the similarity of the resulting nonanalyticities notwithstanding: The similarities
just reflects the fact that the least irrelevant operators have the same scale dimensions in both cases, but they obscure
the fact that they belong to different fixed points. 

In $d=3$ spatial dimensions the least irrelevant operators with respect to the DLF fixed point must have a scale dimension of 2, which leads to the
logarithmic behavior we have found. In this sense $d=3$ represents an upper critical dimension that marks the
boundary between dimensions $d<3$ where the least irrelevant operators represent the leading corrections to the
fixed-point behavior, and dimensions $d>3$ where the leading corrections are analytic. As we have seen in Sec.~\ref{sec:IV},
the prefactors of some of the $T^2\ln T$ and analogous terms vanish in perturbation theory. There is no structural
reason for this that we are aware of. These zeros are likely accidental and an artifact of the low order in the
perturbation theory and/or the loop expansion. One example of such an accidental zero is the absence of a
$T^2\ln T$ contribution to the spin susceptibility of a LFL to second order in perturbation theory in 
$d=3$.\cite{Carneiro_Pethick_1977, Belitz_Kirkpatrick_Vojta_1997} By contrast, the specific-heat coefficient in $d=3$
does have a $T^2\ln T$ contribution,\cite{Chubukov_Maslov_Millis_2006} and an analysis of the spin susceptibility
for generic dimensions yields a $T^{d-1}$ nonanalyticity with a $d$-dependent prefactor that changes sign and is
zero for $d=3$.\cite{us_tbp}

\subsection{Consequences for quantum phase transitions in Dirac metals}
\label{subsec:V.D}

We finally mention that our results for the spin susceptibility of a DFL have important implications for various
magnetic quantum phase transitions in Dirac metals. It is well known that a nonanalytic dependence of the
spin susceptibility on the magnetic field $h$ leads to a corresponding nonanalyticity in the magnetic equation
of state.\cite{Belitz_Kirkpatrick_Vojta_1997, Belitz_Kirkpatrick_Vojta_1999, Brando_et_al_2016a} 
Equation~(\ref{eq:4.7}) implies, at $T=0$ and in a mean-field approximation for the dimensionless magnetization $m$, an
equation of state of the form
\be
t\,m - w\,m^3\ln m + u\,m^3 = h
\label{eq:5.3}
\ee
Here $t$ and $u>0$ are Landau parameters, and $w>0$ due to the sign of the prefactor in Eq.~(\ref{eq:4.7}).\cite{prefactor_sign_footnote}
This implies that any quantum phase transition from a paramagnetic Dirac metal to a homogeneous ferromagnetic
one is necessarily first order. (This applies only at zero temperature, and in the absence of quenched disorder.)
There are, however, other possibilities, such as a transition to a state with modulated magnetic order, with the
wave-number scale of the modulation given by the maximum in the wave-number dependent spin susceptibility.
Analogous conclusions apply to spin-nematic quantum transitions in Dirac metals. Also of interest are transitions
to helimagnetic states with a long pitch wavelength, but our the SI-invariant model studied in the present paper
does not directly apply to those. These implications will be pursued in a separate project.

\acknowledgments

This work was supported by the NSF under grant numbers DMR-1401449 and DMR-1401410. Part of this work was performed at the
Aspen Center for Physics, supported by the NSF under Grant No. PHYS-1066293, and at the Telluride Science Research Center (TSRC). 
We thank George de Coster for discussions.

\bigskip
\appendix

\section{Electron-electron interaction in a Dirac metal}
\label{app:A}

Consider a Coulomb interaction between fluctuations of the particle number density
\be
n(q) = \sum_{\alpha} n^{\alpha}(q) = \sum_{k,\alpha}  \Bigl(\bar\psi^{\alpha}(k),\sigma_0 \psi^{\alpha}(k-q)\Bigr)
\label{eq:A.1}
\ee
Here and in what follows we use an obvious spinor notation: $\psi^{\alpha}(k) = (\psi_{\uparrow}^{\alpha},\psi_{\downarrow}^\alpha)$,
and $\psi(k) =(\psi_{\uparrow}^{+},\psi_{\downarrow}^{+}, \psi_{\uparrow}^-,\psi_{\downarrow}^-)$, and the same four-vector
notation $k = ({\bm k},i\omega_n)$, $q = ({\bm q},i\Omega_n)$ as in Sec.~\ref{sec:III}.
With $v({\bm q})$ the interaction potential, the interaction part of the action can be written
\begin{widetext}
\bea
S_{\text{int}}^{\text c} &=& \frac{-T}{2V} \sum_{q} v({\bm q}) \sum_{k,p} \sum_{\sigma,\sigma' } \sum_{\alpha\alpha'}
     \bar\psi_{\sigma}^{\alpha}(k) \psi_{\sigma}^{\alpha}(k-q)\,\bar\psi_{\sigma'}^{\alpha'}(p-q)\psi_{\sigma'}^{\alpha'}(p)
\nonumber\\
                                  &=& \frac{T}{2V} \sum_{q} v({\bm k}-{\bm p}) \sum_{k,p} \sum_{\sigma,\sigma' } \sum_{\alpha\alpha'}
     \bar\psi_{\sigma}^{\alpha}(k) \psi_{\sigma'}^{\alpha'}(k-q)\,\bar\psi_{\sigma'}^{\alpha'}(p-q)\psi_{\sigma}^{\alpha}(p)
\nonumber\\
                                  &=& \frac{-T}{2V} \sum_{q} v({\bm k}+{\bm p}) \sum_{k,p}\sum_{\sigma,\sigma' } \sum_{\alpha\alpha'}
     \bar\psi_{\sigma}^{\alpha}(-k) \bar\psi_{\sigma'}^{\alpha'}(k-q)\,\psi_{\sigma'}^{\alpha'}(-p-q)\psi_{\sigma}^{\alpha}(p)     
     \label{eq:A.2}
\eea
These are three identical ways to write the interaction. The first two represent the small-angle and large-angle scattering processes
in the particle-hole channel, respectively, from Ref.~\onlinecite{Belitz_Kirkpatrick_Vojta_1997}. The third one represents
scattering in the particle-particle channel, i.e., scattering across the Fermi surface. The latter processes do not contribute to the leading 
singularities we are interested in, and we neglect them. Since we are interested in metals (as defined in Sec.~\ref{sec:I}) 
we can consider the interaction potential $v({\bm q})$ to be screened and hence short ranged.

The singular behavior we are interested in manifests itself at small values of the scattering wave vector ${\bm q}$ (the
`hydrodynamic regime'). If we restrict the sum over this hydrodynamic wave vector to values $\vert{\bm q}\vert < \Lambda$,
with $\Lambda$ a UV cutoff,\cite{cutoff_footnote} we can add the first two identical expressions for $S_{\text{int}}$ in 
Eq.~(\ref{eq:A.2}) to write an effective interaction 
\be
S_{\text{int}} = \frac{-T}{2V} \sum_{q}{}^{'}  \sum_{k,p} \sum_{\sigma,\sigma' }\sum_{\alpha,\alpha'} \left( v({\bm q})
     \bar\psi_{\sigma}^{\alpha}(k) \psi_{\sigma}^{\alpha}(k-q)\,\bar\psi_{\sigma'}^{\alpha'}(p-q)\psi_{\sigma'}^{\alpha'}(p)
     + v({\bm k}-{\bm p}) 
     \bar\psi_{\sigma}^{\alpha}(k) \psi_{\sigma'}^{\alpha'}(k-q)\,\bar\psi_{\sigma'}^{\alpha'}(p-q)\psi_{\sigma}^{\alpha}(p) \right)
     \label{eq:A.3}
\ee
Here the prime on the sum over $q$ indicates the restriction $\vert{\bm q}\vert < \Lambda$. 

The two terms in Eq.~(\ref{eq:A.3}) can be decomposed into a spin singlet and a spin triplet by using the
completeness relation for the Pauli matrices, 
\be
{\bm\sigma}_{\sigma_1\sigma_2}\cdot{\bm\sigma}_{\sigma_3\sigma_4} = 2\delta_{\sigma_1\sigma_4}\delta_{\sigma_2\sigma_3} - \delta_{\sigma_1\sigma_2}\delta_{\sigma_3\sigma_4}\ ,
\label{eq:A.4}
\ee
and the chirality dependence can be written analogously. We obtain
\bea
S_{\text{int}} = \frac{-T}{2V} \sum_{q}{}^{'}  \sum_{k,p} &&\left[ \left(v({\bm q}) - \frac{1}{2}\,v({\bm k}-{\bm p})\right)
     \left(\bar\psi(k)(\pi_0\otimes\sigma_0)\psi(k-q)\right)\left(\bar\psi(p-q)(\pi_0\otimes\sigma_0)\psi(p)\right) \right.
\nonumber\\
   &&-\frac{1}{8}\,v({\bm k}-{\bm p}) \sum_{i=1}^3 
            \left(\bar\psi(k)(\pi_0\otimes\sigma_i)\psi(k-q)\right)\left(\bar\psi(p-q)(\pi_0\otimes\sigma_i)\psi(p)\right)     
\nonumber\\
   &&-\frac{1}{8}\,v({\bm k}-{\bm p}) \sum_{j=1}^3 
            \left(\bar\psi(k)(\pi_j\otimes\sigma_0)\psi(k-q)\right)\left(\bar\psi(p-q)(\pi_j\otimes\sigma_0)\psi(p)\right) 
\nonumber\\
   &&\left. -\frac{1}{8}\,v({\bm k}-{\bm p}) \sum_{i,j=1}^3 
            \left(\bar\psi(k)(\pi_j\otimes\sigma_i)\psi(k-q)\right)\left(\bar\psi(p-q)(\pi_j\otimes\sigma_i)\psi(p)\right) 
             \right]
\label{eq:A.5}             
\eea
In the calculation of any observable, the wave vectors ${\bm k}$ and ${\bm p}$ will be pinned to the Fermi
surface by the poles of the Green functions. One can thus expand $v({\bm k}-{\bm p})$ in a complete set
of functions on the Fermi surface. For our isotropic model this results in an expansion in spherical harmonics,
as in Landau Fermi-liquid theory.\cite{Abrikosov_Gorkov_Dzyaloshinski_1963} For our purposes the $\ell=0$ (s-wave)
term in this expansion suffices, which amounts to replacing $v({\bm k}-{\bm p})$ by its average over the
Fermi surface, and for studying leading the hydrodynamic singularities we can replace $v({\bm q})$ by
$v({\bm q}=0)$. Furthermore, the four terms in Eq.~(\ref{eq:A.5}) are individually invariant under
rotations in spin space and chirality space, and their respective coupling constants will behave differently
under renormalization. We thus obtain Eq.~(\ref{eq:A.5}) with four independent number-valued interaction
amplitudes for the four different terms:
\bea
S_{\text{int}} = \frac{-T}{2V} \sum_{q}{}^{'}  \sum_{k,p} &&\Bigl[ \Gamma_{\text{s}}^{\text{s}}
     \left(\bar\psi(k)(\pi_0\otimes\sigma_0)\psi(k-q)\right)\left(\bar\psi(p-q)(\pi_0\otimes\sigma_0)\psi(p)\right)
\nonumber\\
   &&-\frac{1}{8}\,\Gamma_{\text{t}}^{\text{s}} \sum_{i=1}^3 
            \left(\bar\psi(k)(\pi_0\otimes\sigma_i)\psi(k-q)\right)\left(\bar\psi(p-q)(\pi_0\otimes\sigma_i)\psi(p)\right)     
\nonumber\\
   &&-\frac{1}{8}\,\Gamma_{\text{s}}^{\text{t}} \sum_{j=1}^3 
            \left(\bar\psi(k)(\pi_j\otimes\sigma_0)\psi(k-q)\right)\left(\bar\psi(p-q)(\pi_j\otimes\sigma_0)\psi(p)\right) 
\nonumber\\
   && -\frac{1}{8}\,\Gamma_{\text{t}}^{\text{t}} \sum_{i,j=1}^3 
            \left(\bar\psi(k)(\pi_j\otimes\sigma_i)\psi(k-q)\right)\left(\bar\psi(p-q)(\pi_j\otimes\sigma_i)\psi(p)\right) 
             \Bigr]
\label{eq:A.6}             
\eea
\end{widetext}

Equation~(\ref{eq:A.6}) represents the most general interaction in the s-wave particle-hole channel that is
invariant under rotations in both spin and chirality space. In particular, it is the most general interaction in
that channel for the model defined in Sec.~\ref{sec:II} in the LFL limit, i.e., for a vanishing spin-orbit interaction, $v=0$. 
$v\neq 0$ breaks the rotational invariance in chirality space, and $S_{\text{int}}$ as written in Eq.~(\ref{eq:A.6})
therefore still has a higher symmetry than the single-particle action. If we require invariance only under spatial
inversion, rather than under rotation in chirality space, we see that the four terms in Eq.~(\ref{eq:A.6}) give
rise to six different structures that are all individually invariant under spatial inversion and simultaneous rotations
in spin space and real space and come with six independent interaction amplitudes.\cite{independent_amplitudes_footnote} 
These are given in Eq.~(\ref{eq:2.15}).

\section{Ward identities for a Landau Fermi liquid from rotations in frequency space}
\label{app:B}

Consider noninteracting electrons described by the action $S_0$ in Eq.~(\ref{eq:2.13}) with $v=0$. The partition function is given by
\be
Z = \int D[\bar\psi,\psi]\ e^{S_0[\bar\psi,\psi]}
\label{eq:B.1}
\ee
We augment the action by a source term for products $\bar\psi\psi$,
\bse
\label{eq:B.1.1}
\bea
S_J[\bar\psi,\psi] &=& \sum_{nm} \int d{\bm x}\,d{\bm y} \sum_{\sigma\sigma'}\sum_{\alpha} J^{\alpha}_{nm,\sigma\sigma'}({\bm x},{\bm y})
\nonumber\\
&& \times \bar\psi^{\alpha}_{\sigma}({\bm x},\omega_n)\,\psi^{\alpha}_{\sigma'}({\bm y},\omega_m)
\label{eq:B.1.1a}
\eea
and define a generating functional
\be
Z[J] = \int D[\bar\psi,\psi]\ e^{S_0[\bar\psi,\psi]+S_J[\bar\psi\psi]}
\label{eq:B.1.1b}
\ee
\ese
Now consider transformations of the fermionic fields
\bse
\label{eqs:B.2}
\bea
\psi_{\sigma}^{\alpha}({\bm x},\omega_n) \rightarrow \int d{\bm y} \sum_m \sum_{\sigma'} T^{(i)}_{\substack{  \\nm\\ \sigma\sigma'}}({\bm x},{\bm y})\,\psi_{\sigma'}^{\alpha}({\bm y},\omega_m)\ ,
\nonumber\\
\label{eq:B.2a}\\
\bar\psi_{\sigma}^{\alpha}({\bm x},\omega_n) \rightarrow \int d{\bm y} \sum_m \sum_{\sigma'} T^{(i)*}_{\substack{  \\nm\\ \sigma\sigma'}}({\bm x},{\bm y})\,\bar\psi_{\sigma'}^{\alpha}({\bm y},\omega_m)\ ,
\nonumber\\
\label{eq:B.2b}
\eea
\ese
This defines four different transformations for $i=0,1,2,3$. The $T^{(i)}$ are defined as
\bse
\label{eqs:B.3}
\bea
T^{(i)}_{nm}({\bm x},{\bm y}) &=& \delta_{nm}\,\delta({\bm x}-{\bm y})\,\delta_{\sigma\sigma'} 
\nonumber\\
&& \hskip -10pt +\, t_{nm} \left[\phi({\bm x},{\bm y}) + \phi({\bm y},{\bm x})\right]\left(\sigma^i\right)_{\sigma\sigma'}
\label{eq:B.3a}
\eea
with $\sigma^0$ the $2\times 2$ unit matrix and $\sigma^{1,2,3}$ the Pauli matrices, and
\be
t_{nm} = \delta_{n n_1}\delta_{m n_2} - \delta_{n n_2} \delta_{m n_1} \ .
\label{eq:B.3b}
\ee
\ese
$n_1$ and $n_2 \neq n_1$ are two fixed Matsubara frequency labels, and $\phi({\bm x},{\bm y})$ is a nonlocal mixing angle. 
Note that $\psi$ and $\bar\psi$ transform via $T^{(i)}$ and its complex conjugate $T^{(i)*}$, respectively.

All four of the transformations are unitary with respect to the scalar product of $\bar\psi$ and $\psi$ that is defined by Eq.~(\ref{eq:B.1.1a}) with $J$ replaced by the
unit matrix. The functional integration measure is therefore invariant under the transformations, and we have
\bse
\label{eqs:B.5}
\bea
Z[J] &=& \int D[\bar\psi,\psi]\ e^{S_0 + S_J}
\nonumber\\
       &=& \int D[\bar\psi,\psi]\ e^{S_0 + \delta S_0 + S_J + \delta S_J}
\nonumber\\
       &=& \int D[\bar\psi,\psi]\ e^{S_0 + S_J }\,\left[1 + \delta S_0 + \delta S_J + O(\phi^2)\right]\ .
\nonumber\\
\label{eq:B.5a}
\eea   
or
\be
0 = \int D[\bar\psi,\psi]\ \left(\delta S_0 + \delta S_J\right) e^{S_0 + S_J}\ .
\label{eq:B.5b}
\ee
\ese
Here $\delta S_0$ and $\delta S_J$ are the variations of the action under any of the transformations. If we differentiate Eq.~(\ref{eq:B.5b}) with respect
to $J$ and put $J=0$ we obtain
\bea
\langle \delta S_0\,\bar\psi^{\alpha}_{\sigma}({\bm x},\omega_n)\,\psi^{\alpha}_{\sigma'}({\bm y},\omega_m)\rangle &=& 
      - \langle\delta\bar\psi^{\alpha}_{\sigma}({\bm x},\omega_n)\,\psi^{\alpha}_{\sigma'}({\bm y},\omega_m)\rangle
      \nonumber\\
      && - \langle\bar\psi^{\alpha}_{\sigma}({\bm x},\omega_n)\,\delta\psi^{\alpha}_{\sigma'}({\bm y},\omega_m)\rangle
      \nonumber\\
\label{eq:B.6}
\eea
where $\langle\ldots\rangle$ is an average with respect to the action $S_0$. 

Equation (\ref{eq:B.6}) represents four Ward identities, one for each of the four transformations defined above. Calculating the variations,
and factorizing the four-fermion correlation on the left-hand side of Eq.~(\ref{eq:B.6}) we find that the four identities provide two independent
pieces of information: The transformations with $i=0$ or $i=3$ relate products $G_{n_1\sigma}G_{n_2\sigma}$ to the difference of the same
Green functions, and those with $i=1$ or $i=2$ do the same for $G_{n_1\sigma}G_{n_2,-\sigma}$. The results are equivalent to Eq~(\ref{eq:3.1}).

This derivation emphasizes that the soft modes are the Goldstone modes of a broken rotational symmetry in frequency space that reflects
the difference between retarded and advanced degrees of freedom. The order parameter associated with this broken symmetry is the
spectrum of the Green function, which is nonzero everywhere inside the band. This interpretation of the soft modes as Goldstone modes was first discussed by
Wegner in the context of disordered electron systems.\cite{Wegner_1979}

\section{Symmetry properties of correlation functions}
\label{app:C}

Here we list symmetry properties of convolutions of Green functions that are useful for simplifying the integrals one encounters in
perturbation theory. They can be checked by explicit calculations and are valid only for the leading singular behavior of the convolutions.

Define a convolution
\bse
\label{eqs:C.1}
\be
J^{(3)\alpha\beta}(q,Q) = \int_k f(\hat{\bm k}) F_k^{\alpha\beta} F_{k-q}^{\alpha\beta} F_{k-Q}^{\alpha\beta}
\label{eq:C.1a}
\ee
with $f(\hat{\bm k})$ an arbitrary tensor-valued function of the components of the unit wave vector $\hat{\bm k}$, i.e., 
$f(\hat{\bm k}) = \text{const.}, {\hat k}_i, {\hat k}_i{\hat k}_j$, etc. $J^{(3)}$ scales as
$J^{(3)} \sim 1/q$. The leading singular contributions obey
\be
J^{(3)\alpha\beta}(-q,-Q) = -J^{(3)\alpha\beta}(q,Q)
\label{eq:C.1b}
\ee
\ese
Similarly the convolutions
\bse
\label{eqs:C.2}
\be
{\tilde J}^{(3)\beta}_{\pm\mp\pm}(q,Q) = \sum_k f(\hat{\bm k}) F_k^{\pm\beta} F_{k-q}^{\mp\beta} F_{k-Q}^{\pm\beta}
\label{eq:C.2a}
\ee
obey
\be
{\tilde J}^{(3)\beta}_{+-+}(q,Q) = - {\tilde J}^{(3)\beta}_{-+-}(-q,-Q) 
\label{eq:C.2b}
\ee
\ese

Useful properties of convolutions of four Green functions involve
\bse
\label{eqs:C.3}
\bea
J^{(4)\alpha\beta}_1(q,Q) &=& \int_k f(\hat{\bm k}) F_k^{\alpha\beta} F_k^{\alpha\beta} F_{k-q}^{\alpha\beta} F_{k-Q}^{\alpha\beta}
\label{eq:C.3a}\\
J^{(4)\alpha\beta}_2(q,Q) &=& \int_k f(\hat{\bm k}) F_k^{\alpha\beta} F_{k-q}^{\alpha\beta} F_{k-Q}^{\alpha\beta} F_{k-q-Q}^{\alpha\beta}
\nonumber\\
\label{eq:C.3b}
\eea
which scale as $1/q^2$. One finds, for the leading singular behavior,
\be
J^{(4)\alpha\beta}_2(q,Q) = -2 J^{(4)\alpha\beta}_1(q,Q)
\label{eq:C.3c}
\ee
\ese

Finally, the convolutions
\bse
\label{eqs:C.4}
\bea
{\tilde J}^{(4)\beta}_{1,\pm\pm\mp\pm}(q,Q) &=& \int_k f(\hat{\bm k}) F_k^{\pm\beta} F_k^{\pm\beta} F_{k-q}^{\mp\beta} F_{k-Q}^{\pm\beta}
\label{eq:C.4a}\\
{\tilde J}^{(4)\beta}_{2,\pm\mp\pm\mp}(q,Q) &=& \int_k f(\hat{\bm k}) F_k^{\pm\beta} F_{k-q}^{\mp\beta} F_{k-Q}^{\pm\beta} F_{k-q-Q}^{\mp\beta}
\nonumber\\
\label{eq:C.4b}
\eea
obey
\be
{\tilde J}^{(4)\beta}_{2,+-+-}(q,Q) = -2 {\tilde J}^{(4)\beta}_{1,++-+}(q,Q)
\label{eq:C.4c}
\ee
\ese

These symmetry properties are generalizations of analogous properties in the LFL case.\cite{Belitz_Kirkpatrick_Vojta_1997}


\begin{thebibliography}{59}
\expandafter\ifx\csname natexlab\endcsname\relax\def\natexlab#1{#1}\fi
\expandafter\ifx\csname bibnamefont\endcsname\relax
  \def\bibnamefont#1{#1}\fi
\expandafter\ifx\csname bibfnamefont\endcsname\relax
  \def\bibfnamefont#1{#1}\fi
\expandafter\ifx\csname citenamefont\endcsname\relax
  \def\citenamefont#1{#1}\fi
\expandafter\ifx\csname url\endcsname\relax
  \def\url#1{\texttt{#1}}\fi
\expandafter\ifx\csname urlprefix\endcsname\relax\def\urlprefix{URL }\fi
\providecommand{\bibinfo}[2]{#2}
\providecommand{\eprint}[2][]{\url{#2}}

\bibitem[{\citenamefont{Herring}(1937)}]{Herring_1937}
\bibinfo{author}{\bibfnamefont{C.}~\bibnamefont{Herring}},
  \bibinfo{journal}{Phys. Rev.} \textbf{\bibinfo{volume}{52}},
  \bibinfo{pages}{365} (\bibinfo{year}{1937}).

\bibitem[{\citenamefont{Abrikosov and
  Beneslavskii}(1970)}]{Abrikosov_Beneslavskii_1970}
\bibinfo{author}{\bibfnamefont{A.~A.} \bibnamefont{Abrikosov}}
  \bibnamefont{and} \bibinfo{author}{\bibfnamefont{S.~D.}
  \bibnamefont{Beneslavskii}}, \bibinfo{journal}{Zh. Eksp. Teor. Fiz.}
  \textbf{\bibinfo{volume}{59}}, \bibinfo{pages}{1280} (\bibinfo{year}{1970}),
  \bibinfo{note}{[Sov. Phys. JETP {\bf 32}, 699 (1971)]}.

\bibitem[{\citenamefont{Lifshitz and
  Pitaevskii}(1991)}]{Landau_Lifshitz_IX_1991}
\bibinfo{author}{\bibfnamefont{E.~M.} \bibnamefont{Lifshitz}} \bibnamefont{and}
  \bibinfo{author}{\bibfnamefont{L.~P.} \bibnamefont{Pitaevskii}},
  \emph{\bibinfo{title}{Statistical Physics, Part 2}}
  (\bibinfo{publisher}{Pergamon, Oxford}, \bibinfo{year}{1991}),
  \bibinfo{note}{$\S$ 68}.

\bibitem[{\citenamefont{Wan et~al.}(2011)\citenamefont{Wan, Turner, Vishwanath,
  and Savrasov}}]{Wan_et_al_2011}
\bibinfo{author}{\bibfnamefont{X.}~\bibnamefont{Wan}},
  \bibinfo{author}{\bibfnamefont{A.~M.} \bibnamefont{Turner}},
  \bibinfo{author}{\bibfnamefont{A.}~\bibnamefont{Vishwanath}},
  \bibnamefont{and} \bibinfo{author}{\bibfnamefont{S.~Y.}
  \bibnamefont{Savrasov}}, \bibinfo{journal}{Phys. Rev. B}
  \textbf{\bibinfo{volume}{83}}, \bibinfo{pages}{205101}
  (\bibinfo{year}{2011}).

\bibitem[{\citenamefont{Yang et~al.}(2011)\citenamefont{Yang, Lu, and
  Ran}}]{Yang_Lu_Ran_2011}
\bibinfo{author}{\bibfnamefont{K.}~\bibnamefont{Yang}},
  \bibinfo{author}{\bibfnamefont{Y.}~\bibnamefont{Lu}}, \bibnamefont{and}
  \bibinfo{author}{\bibfnamefont{Y.}~\bibnamefont{Ran}},
  \bibinfo{journal}{Phys. Rev. B} \textbf{\bibinfo{volume}{84}},
  \bibinfo{pages}{075129} (\bibinfo{year}{2011}).

\bibitem[{\citenamefont{Burkov and Balents}(2011)}]{Burkov_Balents_2011}
\bibinfo{author}{\bibfnamefont{A.~A.} \bibnamefont{Burkov}} \bibnamefont{and}
  \bibinfo{author}{\bibfnamefont{L.}~\bibnamefont{Balents}},
  \bibinfo{journal}{Phys. Rev. Lett.} \textbf{\bibinfo{volume}{107}},
  \bibinfo{pages}{127205} (\bibinfo{year}{2011}).

\bibitem[{\citenamefont{Zhang et~al.}(2009)\citenamefont{Zhang, {C-X.~Liu},
  {X-L.~Qi}, Dai, Fang, and {S-C.~Zhang}}}]{Zhang_et_al_2009}
\bibinfo{author}{\bibfnamefont{H.}~\bibnamefont{Zhang}},
  \bibinfo{author}{\bibnamefont{{C-X.~Liu}}},
  \bibinfo{author}{\bibnamefont{{X-L.~Qi}}},
  \bibinfo{author}{\bibfnamefont{X.}~\bibnamefont{Dai}},
  \bibinfo{author}{\bibfnamefont{Z.}~\bibnamefont{Fang}}, \bibnamefont{and}
  \bibinfo{author}{\bibnamefont{{S-C.~Zhang}}}, \bibinfo{journal}{Nature Phys.}
  \textbf{\bibinfo{volume}{5}}, \bibinfo{pages}{438} (\bibinfo{year}{2009}).

\bibitem[{\citenamefont{Belitz et~al.}(2005)\citenamefont{Belitz, Kirkpatrick,
  and Vojta}}]{Belitz_Kirkpatrick_Vojta_2005}
\bibinfo{author}{\bibfnamefont{D.}~\bibnamefont{Belitz}},
  \bibinfo{author}{\bibfnamefont{T.~R.} \bibnamefont{Kirkpatrick}},
  \bibnamefont{and} \bibinfo{author}{\bibfnamefont{T.}~\bibnamefont{Vojta}},
  \bibinfo{journal}{Rev. Mod. Phys.} \textbf{\bibinfo{volume}{77}},
  \bibinfo{pages}{579} (\bibinfo{year}{2005}).

\bibitem[{\citenamefont{Pomeau and Resibois}(1975)}]{Pomeau_Resibois_1975}
\bibinfo{author}{\bibfnamefont{Y.}~\bibnamefont{Pomeau}} \bibnamefont{and}
  \bibinfo{author}{\bibfnamefont{P.}~\bibnamefont{Resibois}},
  \bibinfo{journal}{Phys. Rep.} \textbf{\bibinfo{volume}{2}},
  \bibinfo{pages}{63} (\bibinfo{year}{1975}).

\bibitem[{\citenamefont{Dorfman}(1975)}]{Dorfman_1975}
\bibinfo{author}{\bibfnamefont{J.~R.} \bibnamefont{Dorfman}}, in
  \emph{\bibinfo{booktitle}{{Fundamental Problems in Statistical Mechanics,
  Vol. 3}}}, edited by \bibinfo{editor}{\bibfnamefont{E.~G.~D.}
  \bibnamefont{Cohen}} (\bibinfo{publisher}{North Hollandr},
  \bibinfo{address}{{Amsterdam}}, \bibinfo{year}{1975}), p.
  \bibinfo{pages}{277}.

\bibitem[{\citenamefont{Dorfman et~al.}(1994)\citenamefont{Dorfman,
  Kirkpatrick, and Sengers}}]{Dorfman_Kirkpatrick_Sengers_1994}
\bibinfo{author}{\bibfnamefont{J.~R.} \bibnamefont{Dorfman}},
  \bibinfo{author}{\bibfnamefont{T.~R.} \bibnamefont{Kirkpatrick}},
  \bibnamefont{and} \bibinfo{author}{\bibfnamefont{J.~V.}
  \bibnamefont{Sengers}}, \bibinfo{journal}{Ann. Rev. Phys. Chem.}
  \textbf{\bibinfo{volume}{45}}, \bibinfo{pages}{213} (\bibinfo{year}{1994}).

\bibitem[{\citenamefont{Vaks et~al.}(1967)\citenamefont{Vaks, Larkin, and
  Pikin}}]{Vaks_Larkin_Pikin_1967}
\bibinfo{author}{\bibfnamefont{V.~G.} \bibnamefont{Vaks}},
  \bibinfo{author}{\bibfnamefont{A.~I.} \bibnamefont{Larkin}},
  \bibnamefont{and} \bibinfo{author}{\bibfnamefont{S.~A.} \bibnamefont{Pikin}},
  \bibinfo{journal}{Zh. Eksp. Teor. Fiz.} \textbf{\bibinfo{volume}{53}},
  \bibinfo{pages}{1089} (\bibinfo{year}{1967}), \bibinfo{note}{[Sov. Phys. JETP
  {\bf 26}, 647 (1968)]}.

\bibitem[{\citenamefont{Br{\'e}zin and Wallace}(1973)}]{Brezin_Wallace_1973}
\bibinfo{author}{\bibfnamefont{E.}~\bibnamefont{Br{\'e}zin}} \bibnamefont{and}
  \bibinfo{author}{\bibfnamefont{D.~J.} \bibnamefont{Wallace}},
  \bibinfo{journal}{Phys. Rev. B} \textbf{\bibinfo{volume}{7}},
  \bibinfo{pages}{1967} (\bibinfo{year}{1973}).

\bibitem[{\citenamefont{Lee and Ramakrishnan}(1985)}]{Lee_Ramakrishnan_1985}
\bibinfo{author}{\bibfnamefont{P.~A.} \bibnamefont{Lee}} \bibnamefont{and}
  \bibinfo{author}{\bibfnamefont{T.~V.} \bibnamefont{Ramakrishnan}},
  \bibinfo{journal}{Rev. Mod. Phys.} \textbf{\bibinfo{volume}{57}},
  \bibinfo{pages}{287} (\bibinfo{year}{1985}).

\bibitem[{\citenamefont{Altshuler and Aronov}(1984)}]{Altshuler_Aronov_1984}
\bibinfo{author}{\bibfnamefont{B.~L.} \bibnamefont{Altshuler}}
  \bibnamefont{and} \bibinfo{author}{\bibfnamefont{A.~G.}
  \bibnamefont{Aronov}}, \emph{\bibinfo{title}{Electron-Electron Interactions
  in Disordered Systems}} (\bibinfo{publisher}{North-Holland, Amsterdam},
  \bibinfo{year}{1984}), \bibinfo{note}{edited by M. Pollak and A.~L. Efros}.

\bibitem[{\citenamefont{Brando et~al.}(2016)\citenamefont{Brando, Belitz,
  Grosche, and Kirkpatrick}}]{Brando_et_al_2016a}
\bibinfo{author}{\bibfnamefont{M.}~\bibnamefont{Brando}},
  \bibinfo{author}{\bibfnamefont{D.}~\bibnamefont{Belitz}},
  \bibinfo{author}{\bibfnamefont{F.~M.} \bibnamefont{Grosche}},
  \bibnamefont{and} \bibinfo{author}{\bibfnamefont{T.~R.}
  \bibnamefont{Kirkpatrick}}, \bibinfo{journal}{Rev. Mod. Phys.}
  \textbf{\bibinfo{volume}{88}}, \bibinfo{pages}{025006}
  (\bibinfo{year}{2016}).

\bibitem[{chi()}]{chiralities_footnote}
\bibinfo{note}{The necessity of including both chiralities is consistent with
  the Nielsen-Ninomiya theorem,\cite{Nielsen_Ninomiya_1981, Witten_2015} but it
  is not necessary to invoke this; symmetry arguments\cite{Liu_et_al_2010,
  Burkov_2015} suffice}.

\bibitem[{\citenamefont{Armitage et~al.}(2018)\citenamefont{Armitage, Mele, and
  Vishwanath}}]{Armitage_Mele_Vishwanath_2018}
\bibinfo{author}{\bibfnamefont{N.~P.} \bibnamefont{Armitage}},
  \bibinfo{author}{\bibfnamefont{E.~J.} \bibnamefont{Mele}}, \bibnamefont{and}
  \bibinfo{author}{\bibfnamefont{A.}~\bibnamefont{Vishwanath}},
  \bibinfo{journal}{Rev. Mod. Phys.} \textbf{\bibinfo{volume}{90}},
  \bibinfo{pages}{015001} (\bibinfo{year}{2018}).

\bibitem[{\citenamefont{Kotov et~al.}(2012)\citenamefont{Kotov, Uchoa, Pereira,
  Guinea, and {Castro Neto}}}]{Kotov_et_al_2012}
\bibinfo{author}{\bibfnamefont{V.~N.} \bibnamefont{Kotov}},
  \bibinfo{author}{\bibfnamefont{B.}~\bibnamefont{Uchoa}},
  \bibinfo{author}{\bibfnamefont{V.~M.} \bibnamefont{Pereira}},
  \bibinfo{author}{\bibfnamefont{F.}~\bibnamefont{Guinea}}, \bibnamefont{and}
  \bibinfo{author}{\bibfnamefont{A.~H.} \bibnamefont{{Castro Neto}}},
  \bibinfo{journal}{Rev. Mod. Phys.} \textbf{\bibinfo{volume}{84}},
  \bibinfo{pages}{1067} (\bibinfo{year}{2012}).

\bibitem[{\citenamefont{{C-X. Liu} et~al.}(2010)\citenamefont{{C-X. Liu}, {X-L.
  Qi}, Zhang, Dai, Fang, and {S-C.~Zhang}}}]{Liu_et_al_2010}
\bibinfo{author}{\bibnamefont{{C-X. Liu}}}, \bibinfo{author}{\bibnamefont{{X-L.
  Qi}}}, \bibinfo{author}{\bibfnamefont{H.}~\bibnamefont{Zhang}},
  \bibinfo{author}{\bibfnamefont{X.}~\bibnamefont{Dai}},
  \bibinfo{author}{\bibfnamefont{Z.}~\bibnamefont{Fang}}, \bibnamefont{and}
  \bibinfo{author}{\bibnamefont{{S-C.~Zhang}}}, \bibinfo{journal}{Phys. Rev. B}
  \textbf{\bibinfo{volume}{82}}, \bibinfo{pages}{045122}
  (\bibinfo{year}{2010}).

\bibitem[{lar()}]{large_v_footnote}
\bibinfo{note}{However, this is not to say that the physical behavior of all
  observables is necessarily the same, since for large $v$ there are branches
  of the single-particle spectrum that do not contribute to the Fermi surface.}

\bibitem[{\citenamefont{Mitchell and Fritz}(2015)}]{Mitchell_Fritz_2015}
\bibinfo{author}{\bibfnamefont{A.~K.} \bibnamefont{Mitchell}} \bibnamefont{and}
  \bibinfo{author}{\bibfnamefont{L.}~\bibnamefont{Fritz}},
  \bibinfo{journal}{Phys. Rev. B} \textbf{\bibinfo{volume}{92}},
  \bibinfo{pages}{121109(R)} (\bibinfo{year}{2015}).

\bibitem[{spi()}]{spin_rotational_invariance_footnote}
\bibinfo{note}{This is true only in the absence of a magnetic field. In the
  presence of a field the three spin-triplet interaction amplitudes are no
  longer identical. This effect is not important for our purposes, and we
  ignore it.}

\bibitem[{\citenamefont{Velicky}(1969)}]{Velicky_1969}
\bibinfo{author}{\bibfnamefont{B.}~\bibnamefont{Velicky}},
  \bibinfo{journal}{Phys. Rev.} \textbf{\bibinfo{volume}{184}},
  \bibinfo{pages}{614} (\bibinfo{year}{1969}).

\bibitem[{\citenamefont{Wegner}(1979)}]{Wegner_1979}
\bibinfo{author}{\bibfnamefont{F.}~\bibnamefont{Wegner}}, \bibinfo{journal}{Z.
  Phys. B} \textbf{\bibinfo{volume}{35}}, \bibinfo{pages}{207}
  (\bibinfo{year}{1979}).

\bibitem[{dif()}]{diffuson_footnote}
\bibinfo{note}{These diffusive four-fermion excitations, which are soft even
  though the disorder-averaged Green function itself is massive, are often
  referred to as `diffusons'.}

\bibitem[{\citenamefont{Belitz and
  Kirkpatrick}(1997)}]{Belitz_Kirkpatrick_1997}
\bibinfo{author}{\bibfnamefont{D.}~\bibnamefont{Belitz}} \bibnamefont{and}
  \bibinfo{author}{\bibfnamefont{T.~R.} \bibnamefont{Kirkpatrick}},
  \bibinfo{journal}{Phys. Rev. B} \textbf{\bibinfo{volume}{56}},
  \bibinfo{pages}{6513} (\bibinfo{year}{1997}).

\bibitem[{\citenamefont{Belitz and
  Kirkpatrick}(2012)}]{Belitz_Kirkpatrick_2012a}
\bibinfo{author}{\bibfnamefont{D.}~\bibnamefont{Belitz}} \bibnamefont{and}
  \bibinfo{author}{\bibfnamefont{T.~R.} \bibnamefont{Kirkpatrick}},
  \bibinfo{journal}{Phys. Rev. B} \textbf{\bibinfo{volume}{85}},
  \bibinfo{pages}{125126} (\bibinfo{year}{2012}).

\bibitem[{\citenamefont{Abrikosov et~al.}(1963)\citenamefont{Abrikosov, Gorkov,
  and Dzyaloshinski}}]{Abrikosov_Gorkov_Dzyaloshinski_1963}
\bibinfo{author}{\bibfnamefont{A.~A.} \bibnamefont{Abrikosov}},
  \bibinfo{author}{\bibfnamefont{L.~P.} \bibnamefont{Gorkov}},
  \bibnamefont{and} \bibinfo{author}{\bibfnamefont{I.~E.}
  \bibnamefont{Dzyaloshinski}}, \emph{\bibinfo{title}{Methods of Quantum Field
  Theory in Statistical Physics}} (\bibinfo{publisher}{Dover, New York},
  \bibinfo{year}{1963}).

\bibitem[{Fer()}]{Fermi_surface_footnote}
\bibinfo{note}{We note that the AGD approximation relies on the existence of a
  Fermi surface, which is consistent with our definition of a Dirac metal. We
  also note that in general there is more than one Fermi surface. However, for
  small $h$ the values of $\vF$ and $\NF$ for the different Fermi surfaces are
  close to one another in the limits of both small and large values of $v$, and
  we ignore this complication.}

\bibitem[{\citenamefont{Belitz et~al.}(2002)\citenamefont{Belitz, Kirkpatrick,
  and Vojta}}]{Belitz_Kirkpatrick_Vojta_2002}
\bibinfo{author}{\bibfnamefont{D.}~\bibnamefont{Belitz}},
  \bibinfo{author}{\bibfnamefont{T.~R.} \bibnamefont{Kirkpatrick}},
  \bibnamefont{and} \bibinfo{author}{\bibfnamefont{T.}~\bibnamefont{Vojta}},
  \bibinfo{journal}{Phys. Rev. B} \textbf{\bibinfo{volume}{65}},
  \bibinfo{pages}{165112} (\bibinfo{year}{2002}).

\bibitem[{\citenamefont{Belitz and
  Kirkpatrick}(2014)}]{Belitz_Kirkpatrick_2014}
\bibinfo{author}{\bibfnamefont{D.}~\bibnamefont{Belitz}} \bibnamefont{and}
  \bibinfo{author}{\bibfnamefont{T.~R.} \bibnamefont{Kirkpatrick}},
  \bibinfo{journal}{Phys. Rev. B} \textbf{\bibinfo{volume}{89}},
  \bibinfo{pages}{035130} (\bibinfo{year}{2014}).

\bibitem[{\citenamefont{Belitz et~al.}(1997)\citenamefont{Belitz, Kirkpatrick,
  and Vojta}}]{Belitz_Kirkpatrick_Vojta_1997}
\bibinfo{author}{\bibfnamefont{D.}~\bibnamefont{Belitz}},
  \bibinfo{author}{\bibfnamefont{T.~R.} \bibnamefont{Kirkpatrick}},
  \bibnamefont{and} \bibinfo{author}{\bibfnamefont{T.}~\bibnamefont{Vojta}},
  \bibinfo{journal}{Phys. Rev. B} \textbf{\bibinfo{volume}{55}},
  \bibinfo{pages}{9452} (\bibinfo{year}{1997}).

\bibitem[{\citenamefont{Betouras et~al.}(2005)\citenamefont{Betouras, Efremov,
  and Chubukov}}]{Betouras_Efremov_Chubukov_2005}
\bibinfo{author}{\bibfnamefont{J.}~\bibnamefont{Betouras}},
  \bibinfo{author}{\bibfnamefont{D.}~\bibnamefont{Efremov}}, \bibnamefont{and}
  \bibinfo{author}{\bibfnamefont{A.}~\bibnamefont{Chubukov}},
  \bibinfo{journal}{Phys. Rev. B} \textbf{\bibinfo{volume}{72}},
  \bibinfo{pages}{115112} (\bibinfo{year}{2005}).

\bibitem[{\citenamefont{Carneiro and Pethick}(1977)}]{Carneiro_Pethick_1977}
\bibinfo{author}{\bibfnamefont{G.~M.} \bibnamefont{Carneiro}} \bibnamefont{and}
  \bibinfo{author}{\bibfnamefont{C.~J.} \bibnamefont{Pethick}},
  \bibinfo{journal}{Phys. Rev. B} \textbf{\bibinfo{volume}{16}},
  \bibinfo{pages}{1933} (\bibinfo{year}{1977}).

\bibitem[{\citenamefont{Belitz and
  Kirkpatrick}(1994)}]{Belitz_Kirkpatrick_1994}
\bibinfo{author}{\bibfnamefont{D.}~\bibnamefont{Belitz}} \bibnamefont{and}
  \bibinfo{author}{\bibfnamefont{T.~R.} \bibnamefont{Kirkpatrick}},
  \bibinfo{journal}{Rev. Mod. Phys.} \textbf{\bibinfo{volume}{66}},
  \bibinfo{pages}{261} (\bibinfo{year}{1994}).

\bibitem[{\citenamefont{Chubukov et~al.}(2006)\citenamefont{Chubukov, Maslov,
  and Millis}}]{Chubukov_Maslov_Millis_2006}
\bibinfo{author}{\bibfnamefont{A.}~\bibnamefont{Chubukov}},
  \bibinfo{author}{\bibfnamefont{D.}~\bibnamefont{Maslov}}, \bibnamefont{and}
  \bibinfo{author}{\bibfnamefont{A.~J.} \bibnamefont{Millis}},
  \bibinfo{journal}{Phys. Rev. B} \textbf{\bibinfo{volume}{73}},
  \bibinfo{pages}{045128} (\bibinfo{year}{2006}).

\bibitem[{Zer()}]{Zero-gap_footnote}
\bibinfo{note}{We note that there in certain classes of materials the crystal
  lattice symmetry enforces $\Delta=0$, in which case neglecting $\Delta$ does
  not constitute an approximation, see Ref.~\onlinecite{Yang_Nagaosa_2014}.}

\bibitem[{two()}]{two-loop_footnote}
\bibinfo{note}{These diagrams represent the contributions to $\delta\chisL$
  that contain one integral over a hydrodynamic frequency-momentum. In terms of
  integrals, they represent products of convolutions of Green functions, see
  Eqs.~(\ref{eqs:4.2}). There are others, which are also of second order in the
  interaction, that represent one convolution of six Green functions that
  depends on two hydrodynamic frequency-momenta. These additional integrals
  renormalize the prefactors of the nonanalyticies that result from the
  diagrams in Fig.~\ref{fig:4}. In a field-theoretic
  formulation\cite{Belitz_Kirkpatrick_2012a} they correspond to two-loop terms,
  whereas the ones in Fig.~\ref{fig:4} correspond to one-loop terms. The
  additional diagrams thus belong to a qualitatively different class since they
  are of higher order in a loop expansion. Therefore, they cannot cancel the
  contributions from the one-loop terms and we ignore them. A detailed
  discussion of this issue will be given in Ref.~\onlinecite{us_tbp}.}

\bibitem[{hig()}]{higher_angular_momenta_footnote}
\bibinfo{note}{More precisely, there is no logarithmic contribution for $d=3$
  from integrals of this type with angular factors $(\hat{\bm q}\cdot\hat{\bm
  k})^{\ell}$ with $\ell\geq 2$. This implies that the same is true for the
  corresponding nonanalytic contributions from Fermi-liquid parameters with
  angular momentum $\ell\geq 2$ in a Landau Fermi liquid.}

\bibitem[{\citenamefont{Ashrafi et~al.}(2013)\citenamefont{Ashrafi, Rashba, and
  Maslov}}]{Ashrafi_Rashba_Maslov_2013}
\bibinfo{author}{\bibfnamefont{A.}~\bibnamefont{Ashrafi}},
  \bibinfo{author}{\bibfnamefont{E.~I.} \bibnamefont{Rashba}},
  \bibnamefont{and} \bibinfo{author}{\bibfnamefont{D.}~\bibnamefont{Maslov}},
  \bibinfo{journal}{Phys. Rev. B} \textbf{\bibinfo{volume}{88}},
  \bibinfo{pages}{075115} (\bibinfo{year}{2013}).

\bibitem[{\citenamefont{Lundgren and Maciejko}(2015)}]{Lundgren_Maciejko_2015}
\bibinfo{author}{\bibfnamefont{R.}~\bibnamefont{Lundgren}} \bibnamefont{and}
  \bibinfo{author}{\bibfnamefont{J.}~\bibnamefont{Maciejko}},
  \bibinfo{journal}{Phys. Rev. Lett.} \textbf{\bibinfo{volume}{115}},
  \bibinfo{pages}{066401} (\bibinfo{year}{2015}).

\bibitem[{\citenamefont{Zak et~al.}(2010{\natexlab{a}})\citenamefont{Zak,
  Maslov, and Loss}}]{Zak_Maslov_Loss_2010}
\bibinfo{author}{\bibfnamefont{R.~A.} \bibnamefont{Zak}},
  \bibinfo{author}{\bibfnamefont{D.~L.} \bibnamefont{Maslov}},
  \bibnamefont{and} \bibinfo{author}{\bibfnamefont{D.}~\bibnamefont{Loss}},
  \bibinfo{journal}{Phys. Rev. B} \textbf{\bibinfo{volume}{82}},
  \bibinfo{pages}{115415} (\bibinfo{year}{2010}{\natexlab{a}}).

\bibitem[{\citenamefont{Zak et~al.}(2010{\natexlab{b}})\citenamefont{Zak,
  Maslov, and Loss}}]{Zak_Maslov_Loss_2012}
\bibinfo{author}{\bibfnamefont{R.~A.} \bibnamefont{Zak}},
  \bibinfo{author}{\bibfnamefont{D.~L.} \bibnamefont{Maslov}},
  \bibnamefont{and} \bibinfo{author}{\bibfnamefont{D.}~\bibnamefont{Loss}},
  \bibinfo{journal}{Phys. Rev. B} \textbf{\bibinfo{volume}{85}},
  \bibinfo{pages}{115424} (\bibinfo{year}{2010}{\natexlab{b}}).

\bibitem[{\citenamefont{Pines and Nozi{\`e}res}(1989)}]{Pines_Nozieres_1989}
\bibinfo{author}{\bibfnamefont{D.}~\bibnamefont{Pines}} \bibnamefont{and}
  \bibinfo{author}{\bibfnamefont{P.}~\bibnamefont{Nozi{\`e}res}},
  \emph{\bibinfo{title}{The Theory of Quantum Liquids}}
  (\bibinfo{publisher}{Addison-Wesley, Redwood City, CA},
  \bibinfo{year}{1989}).

\bibitem[{\citenamefont{Ma}(1976)}]{Ma_1976}
\bibinfo{author}{\bibfnamefont{S.-K.} \bibnamefont{Ma}},
  \emph{\bibinfo{title}{Modern Theory of Critical Phenomena}}
  (\bibinfo{publisher}{Benjamin, Reading, MA}, \bibinfo{year}{1976}).

\bibitem[{\citenamefont{Shankar}(1994)}]{Shankar_1994}
\bibinfo{author}{\bibfnamefont{R.}~\bibnamefont{Shankar}},
  \bibinfo{journal}{Rev. Mod. Phys.} \textbf{\bibinfo{volume}{66}},
  \bibinfo{pages}{129} (\bibinfo{year}{1994}).

\bibitem[{us_()}]{us_tbp}
\bibinfo{note}{D. Belitz and T.~R. Kirkpatrick, unpublished results.}

\bibitem[{\citenamefont{Belitz et~al.}(1999)\citenamefont{Belitz, Kirkpatrick,
  and Vojta}}]{Belitz_Kirkpatrick_Vojta_1999}
\bibinfo{author}{\bibfnamefont{D.}~\bibnamefont{Belitz}},
  \bibinfo{author}{\bibfnamefont{T.~R.} \bibnamefont{Kirkpatrick}},
  \bibnamefont{and} \bibinfo{author}{\bibfnamefont{T.}~\bibnamefont{Vojta}},
  \bibinfo{journal}{Phys. Rev. Lett.} \textbf{\bibinfo{volume}{82}},
  \bibinfo{pages}{4707} (\bibinfo{year}{1999}).

\bibitem[{pre()}]{prefactor_sign_footnote}
\bibinfo{note}{Purely within perturbation theory, one cannot exclude the
  possibility that the sign of the prefactor changes at higher orders in the
  interaction. However, general arguments strongly suggest that the sign is
  universal and correctly given by the lowest order in perturbation theory, see
  Ref.~\onlinecite{Brando_et_al_2016a}.}

\bibitem[{cut()}]{cutoff_footnote}
\bibinfo{note}{The cutoff $\Lambda$ must be small compared to the atomic
  wave-number scale, but is otherwise arbitrary. The leading singular behavior
  we are interested in will not depend on it.}

\bibitem[{ind()}]{independent_amplitudes_footnote}
\bibinfo{note}{The number of independent interaction amplitudes for related
  models with a point-like interaction has been investigated in
  Refs.~\onlinecite{Maciejko_Nandkishore_2014, Roy_DasSarma_2016,
  Roy_Foster_2018}. Note, however, that a point-like interaction will
  necessarily acquire a momentum dependence under renormalization and therefore
  can only be a starting point.}

\bibitem[{\citenamefont{Nielsen and Ninomiya}(1981)}]{Nielsen_Ninomiya_1981}
\bibinfo{author}{\bibfnamefont{H.~B.} \bibnamefont{Nielsen}} \bibnamefont{and}
  \bibinfo{author}{\bibfnamefont{M.}~\bibnamefont{Ninomiya}},
  \bibinfo{journal}{Phys. Lett.} \textbf{\bibinfo{volume}{105B}},
  \bibinfo{pages}{219} (\bibinfo{year}{1981}).

\bibitem[{\citenamefont{Witten}()}]{Witten_2015}
\bibinfo{author}{\bibfnamefont{E.}~\bibnamefont{Witten}},
  \eprint{arXiv:1510.08698}.

\bibitem[{\citenamefont{Burkov}(2015)}]{Burkov_2015}
\bibinfo{author}{\bibfnamefont{A.~A.} \bibnamefont{Burkov}},
  \bibinfo{journal}{J. Phys.: Condens. Matter} \textbf{\bibinfo{volume}{27}},
  \bibinfo{pages}{113201} (\bibinfo{year}{2015}).

\bibitem[{\citenamefont{Yang and Nagaosa}(2014)}]{Yang_Nagaosa_2014}
\bibinfo{author}{\bibfnamefont{B.}~\bibnamefont{Yang}} \bibnamefont{and}
  \bibinfo{author}{\bibfnamefont{N.}~\bibnamefont{Nagaosa}},
  \bibinfo{journal}{Nature Commun.} \textbf{\bibinfo{volume}{5}},
  \bibinfo{pages}{4898} (\bibinfo{year}{2014}).

\bibitem[{\citenamefont{Maciejko and
  Nandkishore}(2014)}]{Maciejko_Nandkishore_2014}
\bibinfo{author}{\bibfnamefont{J.}~\bibnamefont{Maciejko}} \bibnamefont{and}
  \bibinfo{author}{\bibfnamefont{R.}~\bibnamefont{Nandkishore}},
  \bibinfo{journal}{Phys. Rev. B} \textbf{\bibinfo{volume}{90}},
  \bibinfo{pages}{035126} (\bibinfo{year}{2014}).

\bibitem[{\citenamefont{Roy and {Das Sarma}}(2016)}]{Roy_DasSarma_2016}
\bibinfo{author}{\bibfnamefont{B.}~\bibnamefont{Roy}} \bibnamefont{and}
  \bibinfo{author}{\bibfnamefont{S.}~\bibnamefont{{Das Sarma}}},
  \bibinfo{journal}{Phys. Rev. B} \textbf{\bibinfo{volume}{94}},
  \bibinfo{pages}{115137} (\bibinfo{year}{2016}).

\bibitem[{\citenamefont{Ray and Foster}(2018)}]{Roy_Foster_2018}
\bibinfo{author}{\bibfnamefont{B.}~\bibnamefont{Ray}} \bibnamefont{and}
  \bibinfo{author}{\bibfnamefont{M.~S.} \bibnamefont{Foster}},
  \bibinfo{journal}{Phys. Rev. X} \textbf{\bibinfo{volume}{8}},
  \bibinfo{pages}{011049} (\bibinfo{year}{2018}).

\end{thebibliography}

\end{document}